\documentclass[12pt]{article}
\usepackage{jheppub}
\usepackage{subcaption}
\newcommand\Tr{\mathrm{Tr}}
\newcommand\bC{\mathbb{C}}
\newcommand\bR{\mathbb{R}}
\newcommand\bZ{\mathbb{Z}}

\newcommand\aA{\mathcal{A}}
\newcommand\aB{\mathcal{B}}
\newcommand\aC{\mathcal{C}}

\title{Twisted M2 brane holography and sphere correlation functions}
\author[a]{Davide Gaiotto}
\author[a]{, Jacob Abajian}
\affiliation[a]{Perimeter Institute for Theoretical Physics, Waterloo, Ontario, Canada N2L 2Y5}
\abstract{We define and compute algebraically a ``perturbative part'' of protected sphere correlation functions in the M2 brane SCFTs. These correlation functions are expected to have 
a holographic description in terms of twisted, $\Omega$-deformed M-theory. We uncover a hidden perturbative triality symmetry which supports this conjecture. We also discuss some variants of the setup, involving
M2 branes at $A_k$ singularities and D3 branes with a transverse compact direction.}
\begin{document}
\maketitle

\section{Introduction and plan of the paper}

The objective of this paper is to study potential examples of {\it twisted holography}, in the sense of \cite{Mezei:2017kmw,Costello:2017fbo,Costello:2018zrm,Ishtiaque:2018str}. 
All our examples will take the form of some collection of protected SCFT correlation functions 
encoded in a topological quantum-mechanical system \cite{Gaiotto:2010be,Dimofte:2011py,Beem:2016cbd,Dedushenko:2016jxl,Dedushenko:2017avn}. 
We conjecture them to be holographically dual to twisted M-theory \cite{Costello:2016nkh,Gaiotto:2019wcc} on appropriate backgrounds.

In all of the examples, we will identify hidden structures in the SCFT correlation functions which support the conjecture. We will leave
detailed calculation on the twisted M-theory side to future work. 

Here we reserve the term ``twisted M-theory'' to the five-dimensional holomorphic-topological theory which describes topologically twisted M-theory on an 
$\Omega$-deformed $\bC_{\epsilon_1} \times \bC_{\epsilon_2} \times \bC_{\epsilon_3}$ background \cite{Costello:2016nkh}. 
This theory has a triality symmetry \cite{Gaiotto:2019wcc} which permutes the $\Omega$ deformation parameters $\epsilon_i$. 
We will find an analogous triality symmetry emerging in a very non-trivial way in the protected SCFT correlation functions.

Our main example are the protected sphere correlation functions for the three-dimensional ${\cal N}=8$ ``M2 brane'' SCFT, i.e. the SCFT which appears at low energy on a stack of $N$ $M2$ branes in flat space. We study the correlation functions in the UV description of the SCFT provided by the 
${\cal N}=4$ ADHM gauge theory, i.e. the D2-D6 worldvolume SQFT. Because this theory is self-mirror, we can compute the correlation functions either as ``Higgs branch'' correlation functions or as ``Coulomb branch'' correlation functions. 

Adopting some tricks from the study of the sphere partition function \cite{Marino:2011eh}, we define a grand canonical version of the correlation functions and take a careful large $N$ limit. We conjecture a decomposition of the protected correlation functions into a ``perturbative'' and a ``non-perturbative'' pieces. The perturbative piece manifests a hidden triality invariance, broken by the non-perturbative piece. We conjecture a concise, purely algebraic characterization of the perturbative piece. 

The perturbative piece of the protected correlation functions has the correct structure to be holographically dual to a perturbative twisted M-theory background, which should be a deformation of $S^1 \times \bC \times \bC$. We conduct extensive numerical and algebraic tests of the conjectures. 

We also consider some other examples:
\begin{itemize}
\item The Higgs branch sphere correlation functions for the three-dimensional ${\cal N}=4$ SCFT associated to M2 branes at an $A_1$ singularity. 
We push the analysis as far as for the previous case. The conjectural dual background is a perturbative deformation of $S^1 \times \frac{\bC\times \bC}{\mathbb{Z}_2}$.
\item The Higgs branch sphere correlation functions for the three-dimensional ${\cal N}=4$ SCFT associated to M2 branes at an $A_n$ singularity. 
We only do a partial analysis. The conjectural dual background is a perturbative deformation of $S^1 \times \frac{\bC\times \bC}{\mathbb{Z}_{n+1}}$.
\item The line defect junction Schur indices for the four-dimensional ${\cal N}=4$ SYM with $U(N)$ gauge group. These are natural 4d analogue of Coulomb branch correlation functions, except that they involve BPS line defects wrapping a compact circle in the geometry. We define a somewhat peculiar grand canonical 
version of the correlation functions. Concrete examples of grand canonical correlation functions manifest exact triality invariance up to an overall normalization and some analytic subtleties, without any need of a perturbative expansion. The conjectural dual background is a perturbative deformation of $S^1 \times \bC^* \times \bC^*$.
\end{itemize}  

\section{Protected correlation functions for M2 branes}

The low energy super-conformal field theory residing on the world-volume of $N$ M2 branes is of considerable theoretical interest. 
At large $N$, it gives the best understood example of an holographic duality which is {\it not} based on a 't Hooft expansion, as the expected gravitational dual
is given by M-theory on an $AdS_4 \times S^7$ background \cite{Maldacena:1997re}. 

There is a particularly interesting collection of protected correlation functions of local operators on a three-sphere which are exactly computable via localization \cite{Dedushenko:2016jxl,Dedushenko:2017avn}. 
These correlation functions played a crucial role in a recent, strikingly successful conformal bootstrap analysis \cite{Agmon:2019imm}. Furthermore, it has been proposed \cite{Mezei:2017kmw} 
that the correlation functions should be holographically dual to an analogous protected sector of M-theory on $AdS_4 \times S^7$, giving a 
notable example of ``twisted holography''. 

The protected sphere correlation functions are computed by a topological $U(N)$-gauged matrix quantum mechanics, with a schematic action 
\begin{equation}
\frac{1}{\epsilon_1} \int_{S^1} \Tr \left[\epsilon_2 A_t + X D_t Y + J D_t I \right] dt
\end{equation}
for adjoint fields $X,Y$ and (anti)fundamental fields $I,J$. The correlation functions are computed with anti-periodic boundary conditions for the fields on the circle.  
  
Intuitively, the quantum mechanics describes the supersymmetric motion of the M2 branes along four of the eight transverse directions. 
The basic observables 
\begin{equation}
O_{l m} = \mathrm{STr}\, X^{l+m} Y^{l-m}
\end{equation}
deform the algebra of holomorphic functions of the transverse positions of the $N$ M2 branes in $\bC\times \bC$. 

The main claim of  \cite{Mezei:2017kmw} is that the topological quantum mechanics should have a 
two-dimensional holographic dual description encoding the corresponding protected sector of M-theory on $AdS_4 \times S^7$. 
The two-dimensional theory was presented as a 2d gauge theory with an infinite-dimensional gauge symmetry, 
which is essentially the algebra of complex symplectomorphisms of $\bC\times \bC$. 

The effective action of the ``gravitational'' 2d gauge theory was not determined a-priori, but should be derived order-by-order 
by comparison with the topological QM. In principle, comparison with supergravity localization could then give information about the 
low energy effective action of M-theory.  

We would like to sharpen the proposal by identifying the holographic dual as a five-dimensional holomorphic (symplectic)-topological theory
defined on an $AdS_2 \times S^3$-like background which arises from the localization of the full M-theory background. 
The natural five-dimensional candidate is the $\Omega$-deformed twisted M-theory defined in \cite{Costello:2016nkh}.
This theory is uniquely renormalizable in an appropriate sense, with no adjustable parameters in the effective action
beyond the $\Omega$ deformation parameters. 

The three $\Omega$ deformed factors of the $\bC_{\epsilon_1} \times \bC_{\epsilon_2} \times \bC_{\epsilon_3}$
transverse geometry should correspond to the normal directions to $AdS_2 \times S^3$ in $AdS_4 \times S^7$, in this order. 
In particular, the triality symmetry permuting these factors should hold perturbatively, but may be broken by instanton corrections 
which explore the full transverse geometry.
 
At the local level, this twisted holographic duality is already demonstrated in \cite{Costello:2017fbo}: the OPE of local operators in the topological quantum mechanics 
can be reproduced by perturbative calculations in twisted M-theory on an $\bR \times \bC \times \bC$ background. 
The M2 brane backreaction can be treated perturbatively, as the $N$ dependence of OPE coefficients is polynomial. 
The emergent triality invariance of the OPE was demonstrated in \cite{Gaiotto:2019wcc}. 
\footnote{The local holographic duality can be justified by a simple argument, completely analogous to the argument given in \cite{Costello:2018zrm} for 
D3 brane twisted holography. The argument involves the topological twist and $\Omega$ deformation of the conventional 
Maldacena argument \cite{Maldacena:1997re} for holography, where the world-volume theory of a stack of D-branes in flat space 
becomes dual to the near-horizon limit of the back-reacted geometry. 

We can start from $N$ M2 branes in flat space and apply the deformation. The bulk M-theory becomes Costello's 5d holomorphic-topological 
theory defined on $\bR \times \bC^2$. The M2 brane world-volume theory becomes precisely the auxiliary 1d quantum mechanical system discussed above.
The topological quantum mechanics is coupled in a unique gauge-invariant way to the bulk twisted M-theory \cite{Costello:2016nkh}. 
We naturally deduce that the 1d quantum mechanics should be dual to Costello's theory on whatever 5d background is produced by the M2 branes back-reaction. }

Our objective is to study the full correlation functions of the system, where the topological direction is compactified to a circle. This introduces two new phenomena: 
\begin{itemize}
\item The $N$-dependence of correlation functions is much richer and definitely not polynomial. A careful analysis is required to disentangle the systematic
large $N$ expansion of the correlation functions.  
\item The topological quantum mechanics is not an {\it absolute} theory: there is a non-trivial space of solution of OPE Ward identities, analogous to the space of conformal blocks of a two-dimensional 
chiral algebra. The protected sphere correlation functions of the physical theory give a very specific element of this space of solutions.  
\end{itemize}

One of the most exciting aspects of twisted holography is the possibility to study in an exactly solvable model aspects of quantum gravity such as sums over semiclassical saddles with different geometry. 
The precise holographic interplay between the space of possible solutions of Ward identities and the sum over geometries is not currently understood. 
At the very least, it should select geometries of the twisted gravity theory which can be extended to geometries for the underlying physical 
gravity theory. 

As a preparation to a full holographic analysis, we will accomplish two objectives on the field theory side:
\begin{itemize}
\item We will disentangle the $N$ dependence of correlation functions and identify a ``perturbative part'' which may match perturbative holographic calculations around a dominant 
semiclassical saddle. The perturbative part enjoys the emergent triality symmetry which is expected from the twisted M-theory side.  
\item We will characterize the full space of solutions of the OPE Ward identities and identify a set of conjectural quadratic constraints which uniquely characterize the perturbative 
part of the physical correlation functions in a purely algebraic way. We will test the conjecture both numerically and analytically. 
\end{itemize}
We expect the quadratic constraints, somewhat analogous to the ``string equation'' in topological gravity, to play an important role in 
a direct proof of the twisted holography correspondence. 

\subsection{The BPS algebra}
The M2 brane SCFT has a variety of different gauge theory UV descriptions. We focus on the description which arises from worldvolume theory of $N$ D2 branes in the presence of a single D6 brane, 
i.e. a ${\cal N}=4$ $U(N)$ gauge theory coupled to an adjoint hypermultiplet and a single fundamental hypermultiplet. \footnote{The localization analysis of protected sphere correlation functions is not currently available in 
other descriptions, such as the ABJM theory.}

The protected sphere correlation functions of the M2 brane theory can be identified with the protected Higgs branch correlation functions of the ${\cal N}=4$ theory \cite{Dedushenko:2016jxl}. 
Alternatively, they can be identified with the protected Coulomb branch correlation functions of the same theory \cite{Dedushenko:2017avn}. The two descriptions are isomorphic, but the 
isomorphism is very non-trivial. The Higgs branch presentation only involves polynomials in the elementary fields and preserves the most symmetry. The Coulomb branch presentation  
involves disorder (monopole) operators, but reveals a hidden commutative subalgebra with useful properties. We refer to \cite{Costello:2017fbo,Gaiotto:2019wcc} for a detailed discussion and only review here the results we need for calculations.

\subsubsection{Higgs branch presentation}
The ``quantum'' Higgs branch algebra $\aA_{N}$ 
is defined as a quantum Hamiltonian reduction \cite{Yagi:2014toa,Bullimore:2015lsa}. The operators in the algebra are gauge-invariant polynomials in adjoint elementary fields $(X,Y)$ and (anti)fundamental $(I,J)$. 
The elementary fields have non-trivial commutation relations 
\begin{equation}
[X^a_b, Y^c_d]=\epsilon_1 \delta^a_d\delta^c_b \qquad \qquad [J^b, I_a]=\epsilon_1 \delta^b_a
\end{equation}
and one quotients by the ideal generated by the F-term relation 
\begin{equation}
X^a_c Y^c_b - X^c_b Y^a_c + I_b J^a = \epsilon_2 \delta^a_b
\end{equation}
i.e. the relation can be assumed to hold when placed at the very left (or right) of an operator. 

The algebra $\aA_{N}$ has an $SU(2)$ global symmetry rotating $(X,Y)$ as a doublet. This is an inner 
automorphism of the algebra, generated by 
\begin{equation}
\frac{1}{\epsilon_1} \Tr X^2 \qquad \qquad \frac{1}{\epsilon_1}\Tr Y^2\qquad \qquad \frac{1}{2\epsilon_1} \Tr (XY + YX)
\end{equation}

With the help of the commutation relations and F-term relation, every gauge-invariant operator can be simplified to a polynomial in the elementary symmetrized traces 
\begin{equation}
O_{lm} = \mathrm{STr} X^{l+m} Y^{l-m}
\end{equation}
This claim is not immediately obvious. One can define a collection of moves which applied recursively will lead to the desired result: 
\begin{enumerate}
\item We can apply commutation relations until the operator ordering agrees to the ordering of gauge contractions, so that we have a polynomial in expressions of the form 
$\mathrm{Tr} P(X,Y)$ or $I P(X,Y) J$ where $P(X,Y)$ is some sequence of $X$ and $Y$ fields. Each commutation produces extra terms with fewer symbols, to be simplified recursively. 
\item We can use the F-term relation to reorder the $X$ and $Y$ fields in each sequence, so that we have a polynomial in expressions of the form 
$\mathrm{Tr} S(X,Y)$ or $I S(X,Y) J$ where $S(X,Y)$ is a symmetrized sequence of $X$ and $Y$ fields. Each application of the F-term relations produces extra terms with fewer $X$ and $Y$ symbols, to be simplified recursively. 
\item We can use the F-term relation to map $I S(X,Y) J$ to a polynomial in $\mathrm{Tr} S'(X,Y)$ operators with the same number of or fewer $X$ and $Y$ symbols, to be simplified recursively. Indeed,  if we ignore operator ordering $\Tr [X,Y] S(X,Y)$ vanishes.\footnote{This can be shown e.g. by an $SU(2)$ rotation to $\Tr [X,Y]X^n = \Tr X Y X^n - \Tr Y X^{n+1}=0$.}
\end{enumerate}
The operators $O_{l,-l}, \cdots O_{l,l}$ form an irreducible representation of the $SU(2)$ global symmetry rotating $(X,Y)$ as a doublet. 

If we only use the above transformations to reduce a gauge-invariant operator, such as a commutator $[O_{lm}, O_{l'm'}]$, 
to polynomials in symmetrized traces, the rank $N$ only enters the calculation as the value of $O_{0,0} = \mathrm{Tr} \,1$. 
Following \cite{Costello:2017fbo}, we define an universal algebra $\aA$ with generators $O_{lm}$ and commutation 
relations 
\begin{equation}
[O_{lm}, O_{l'm'}] = \cdots
\end{equation} 
computed by a recursive application of the rules above, with $N$ left arbitrary. 

For any specific value of $N$, the $O_{lm}$ generators will satisfy further polynomial constraints due to the trace relations. 
For example, for $N=1$ one has $O_{lm} O_{l'm'} = O_{l+l',m+m'}$. 
These constraints can be thought of as an algebra morphism $\aA \to \aA_{N}$. The universal algebra $\aA$ will play an important role in our large $N$ analysis. 

In the following, we will find it useful to consider a slightly different normalization and labelling of the basic generators: 
\begin{equation}
t_{m,n} = \frac{1}{\epsilon_1} \mathrm{STr} X^{m} Y^{n}
\end{equation}
We can also package together operators belonging to the same irreducible $SU(2)$ representation into a standard generating function: 
\begin{equation}
t_n(u) = \sum_{m=0}^n {n \choose m} u_1^m u_2^{n-m} t_{n,m} 
\end{equation}

In this normalization, the commutation relations are a non-linear deformation 
\begin{equation}
[ t_{a,b}, t_{c,d} ] = (ad-bc)t_{a+c-1,b+d-1} + O(\epsilon_i)
\end{equation} 
of the commutation relations of the Lie algebra $\mathfrak{s}$ of complex Hamiltonian symplectomorphisms of $\bC^2$:
This is the gauge algebra employed in \cite{Mezei:2017kmw} and identified there as area-preserving diffeomorphisms of the two-sphere.
The presentation of $\aA$ as a deformation of the universal enveloping algebra $U(\mathfrak{s})$ will be useful throughout the paper. 

\subsubsection{A concise presentation}
Notably, the commutation relations defining $\aA$ can all be derived recursively from a simple 
generating set \footnote{This set is actually a bit redundant. For example, the last relation is in the $SU(2)$ orbit of 
\begin{equation}
[ t_{3,0}, t_{0,d} ] = 3 d\, t_{2,d-1} + \sigma_2 \frac{d (d - 1) (d - 2)}{4} t_{0,d-3}+ \frac32 \sigma_3 \sum_{m=0}^{d-3} (m + 1)(d - m - 2) t_{0,m} t_{0,d-3-m}
\end{equation}}:
\begin{align}
[ t_{0,0}, t_{c,d} ] &= 0 \cr
[ t_{1,0}, t_{c,d} ] &= d \,t_{c,d-1}\cr
[ t_{0,1}, t_{c,d} ] &= -c\, t_{c-1,d}\cr
[ t_{2,0}, t_{c,d} ] &= 2 d\, t_{c+1,d-1}  \cr
[ t_{1,1}, t_{c,d} ] &= (d-c) \,t_{c,d}  \cr
[ t_{0,2}, t_{c,d} ] &= - 2 c \,t_{c-1,d+1}  \cr
[ t_{3,0}, t_{c,d} ] &= 3 d\, t_{c+2,d-1} + \sigma_2 \frac{d (d - 1) (d - 2)}{4} t_{c,d-3}+ \cr &+
 \frac32 \sigma_3 \sum_{m=0}^{d-3} \sum_{n=0}^c
   \frac{{m + n + 1 \choose n + 1} (n + 1) {d - m + c - n - 2 \choose
     c - n + 1} (c - n + 1)}{{d + c\choose c}} t_{n,m} t_{c-n,d-3-m}
\end{align}
Here we employed some convenient combinations of the $\epsilon_i$ parameters:
\begin{equation}
\sigma_2 \equiv \epsilon_1^2 + \epsilon_1\epsilon_2 + \epsilon_2^2 \qquad \qquad \sigma_3 = -\epsilon_1 \epsilon_2 (\epsilon_1 + \epsilon_2)
\end{equation}
The commutation relations preserve the scaling symmetry which assigns weight $1$ to $\epsilon_i$ and $\frac{n+m}{2}-1$ to $t_{n,m}$. 

In $SU(2)_R$ invariant notation, with $(u,v) = u_1 v_2 - u_2 v_1$, the generating relations become
\begin{align}
[ t_{0}, t_{c}(v) ] &= 0 \cr
[ t_{1}(u), t_{c}(v) ] &= c (u,v)t_{c-1}(v) \cr
[ t_{2}(u), t_{c}(v) ] &= 2 (u,v) u \cdot \partial_v t_c(v) \cr
[ t_{3}(u), t_{c}(v) ] &= \frac{3}{c+1} (u,v) (u \cdot \partial_v)^2 t_{c+1}(v) + \sigma_2 \frac{c (c -1) (c - 2)}{4}  (u,v)^3 t_{c-3}(v) +\cr 
&+\frac32 \sigma_3 (u,v)^3 \sum_{m=0}^{c-3} (m+1)(c-m-2) t_m(v) t_{c-m-3}(v)
\end{align}

Notice that $t_{2,0}$, $t_{1,1}$, $t_{0,2}$ are the generators of infinitesimal $SU(2)_R$ rotations. Commutators with $t_{1,0} = \epsilon_1^{-1} \Tr X$
and $t_{0,1} = \epsilon_1^{-1} \Tr Y$ are also very easy to compute. The only laborious calculation is the reorganization of $[ t_{3,0}, t_{0,n} ]$ into a polynomial in the $t$'s. 
Once that is done, we can reconstruct $[ t_{3,0}, t_{c,d} ]$ by taking commutators with $t_{2,0}$. We refer the 
reader to the Appendices of \cite{Oh:2020hph} for an example of the algebraic manipulations which can be employed to derive the above commutation relations. 

A further $SU(2)_R$ rotation gives us $[t_{2,1}, t_{c,d}]$ and in particular $t_{n+1,0} = n^{-1} [t_{n,0},t_{2,1}]$, which can be used to recursively compute $[t_{n,0},t_{c,d}]$
and then all other commutators. 
 
This presentation of the algebra makes manifest an important hidden property: triality invariance. Define $\epsilon_3 = - \epsilon_1 - \epsilon_2$. Then 
\begin{equation}
\sigma_2 = \frac12 \sum_{i=1}^3 \epsilon_i^2 \qquad \qquad \sigma_3 = \prod_{i=1}^3 \epsilon_i
\end{equation}
and $\aA$ is manifestly invariant under permutations of the $\epsilon_i$! These are identified with the $\Omega$-deformation parameters of the dual twisted M-theory background. 

Triality is broken at finite $N$ by the value of $t_{0,0}= \frac{N}{\epsilon_1}$. \footnote{The algebra $\aA$ is conjecturally equipped with a three-parameter family of truncations 
$\aA_{N_1, N_2, N_3}$ which specialize the central generator $t_{0,0}$ as
\begin{equation}
t_{0,0} = \frac{N_1}{\epsilon_1}+\frac{N_2}{\epsilon_2}+\frac{N_3}{\epsilon_3}
\end{equation}
and should describe protected local operators at the intersection of three mutually orthogonal stacks of M2 branes. } This makes it obvious that triality can at best be a property of
correlation functions in some large $N$ limit. 

\subsubsection{The Coulomb branch presentation}
The ``quantum'' Coulomb branch algebra of a three-dimensional ${\cal N}=4$ gauge theory has a more intricate practical definition \cite{Braverman:2016wma,Bullimore:2015lsa},
mostly due to the fact that it involves monopole operators. It always includes a commutative subalgebra defined by gauge-invariant polynomials in a single adjoint vectormultiplet field. 

Denote as $\aC_N$ the quantum Coulomb branch of a ${\cal N}=4$ $U(N)$ gauge theory coupled to an adjoint hypermultiplet 
and a single fundamental hypermultiplet, with $\epsilon_1$ being the quantization parameter and $\epsilon_2$ the ``quantum mass parameter'' for the adjoint hypermultiplet. 
As this gauge theory is self-mirror, $\aC_N$ must be isomorphic to $\aA_N$ and provide an alternative presentation of the M2 brane protected algebra. 
The isomorphism, though, is far from trivial. The quantum Coulomb branch algebras $\aC_N$ can be identified with certain spherical Cherednik algebras \cite{Kodera:2016faj},
which can also be identified with $\aA_N$.

They $\aC_N$ algebras have a uniform-in-$N$ description as truncation of a shifted $\mathfrak{gl}(1)$ affine Yangian algebra
$\aC$ \cite{Kodera:2016faj,2019arXiv190307734W} \footnote{More precisely, it can be given as a subalgebra 
of the affine Yangian reviewed in \cite{2014arXiv1404.5240T}, with $e^{\mathrm{here}}_n=e^{\mathrm{there}}_n$, $h^{\mathrm{here}}_n=\psi^{\mathrm{there}}_{n+1}$,$f^{\mathrm{here}}_n=f^{\mathrm{there}}_{n+1}$ and $\psi^{\mathrm{there}}_0=0$.}
equipped with algebra morphisms $\aC \to \aC_N$. The algebra $\aC$ is triality-invariant and conjecturally isomorphic to $\aA$ \cite{Gaiotto:2019wcc}.

Conjecturally, we can build the isomorphism as follows. Define recursively 
\begin{equation}
e_n = - \frac12 [ t_{2,2},e_{n-1}] \qquad \qquad f_n =  \frac12 [ t_{2,2},f_{n-1}]
\end{equation}
starting from $e_0 = t_{0,1}$ and $f_0 = t_{1,0}$. 
Then one observes that 
\begin{equation}
[e_n, f_m] = h_{n+m}
\end{equation}
and the $h_{n}$ commute with each other and with $t_{1,1}$. Furthermore, $t_{2,2}$ is a polynomial in the $h_n$ and we can find other polynomials $d_n$ such that 
\begin{equation}
[d_n, e_m] = - n e_{n+m-1} \qquad \qquad [d_n, f_m] = n f_{n+m-1}
\end{equation}
These relations, the explicit relation between $d_n$ and $h_m$'s and several more Serre and quadratic relations define the affine Yangian. 

In the following, we only need to know that the commuting generators $d_n$ exist, are given as the trace of specific polynomials of the adjoint vectormultiplet field
by the Coulomb branch presentation of the affine Yangian and match specific polynomials in the $t_{a,b}$ generators. 

\subsection{Correlation functions as twisted traces}
Protected sphere correlation functions for any $N$ behave as correlation functions of a topological 1d system. We can compute correlation functions of any ordered sequence of operators, 
with a twisted cyclicity relation 
\begin{equation}
\langle O_1 \cdots O_k t_{n,m} \rangle^{(N)} = (-1)^{n+m} \langle t_{n,m} O_1 \cdots O_k  \rangle^{(N)} 
\end{equation}
In other words, the collection of all protected sphere correlation functions gives a {\it twisted trace} on the quantized Higgs branch algebra $\aA_N$. 

We can immediately promote the correlation functions to a twisted trace for the universal algebra $\aA$, without any loss of information. 
Any operator in $\aA$ which vanishes in $\aA_N$ will vanish when inserted in the correlation functions pulled back from $\aA_N$.

The twisted cyclicity relations are the basic OPE Ward identities satisfied by the correlation functions. They are rather constraining. For example, they determine 
correlation functions involving ``odd'' operators in terms of correlation functions involving even operator only, because if $n+m$ is odd
\begin{equation}
\langle O_1 \cdots O_k t_{n,m} \rangle = \frac12 \langle [O_1 \cdots O_k, t_{n,m}]  \rangle  
\end{equation}
It is easy to see that the odd twisted trace relations do not put any constraint on correlation functions containing even operators only.

Even operators, instead, give symmetries of the correlation functions: if $n+m$ is even, we have
\begin{equation}
\langle [O_1 \cdots O_k, t_{n,m}]  \rangle  = 0
\end{equation} 

We have found ample evidence of the following conjecture: the twisted trace relations allow one to express any correlation function as a linear combination of the ``extremal'' correlators
\begin{equation}
\langle t_{2,0}^{\sum_i n_i} \prod_i t_{0,2n_i} \rangle
\end{equation}
and do not impose any further relations on the extremal correlators. Thus the values of extremal correlators parameterize the space of solutions of OPE Ward identities.

The reduction to extremal correlators proceeds recursively by the transformation
\begin{equation}
\langle \cdots t_{n,m} \cdots \rangle \to \langle \cdots t_{n,m} \cdots \rangle  - c \langle [t_{n-1,m+1}, \cdots t_{2,0} \cdots ]\rangle
\end{equation}
where $c$ is a number selected to set to $1$ the coefficient of $\langle \cdots t_{n,m} \cdots \rangle$ in the commutator and $n+m>2$ or $n=m=1$.
The transformation never requires $c$ to depend on $\sigma_i$. In particular, the reduction to the basis of extremal correlators appears 
to be a property of twisted traces on $U(\mathfrak{s})$ which is inherited by $\aA$.

For each $N$, the actual protected correlation functions will produce a specific solution of the twisted trace relations.
As the space of solutions is linear, any linear combination of correlation functions $\langle O_1 \cdots O_k \rangle^{(N)}$ for different values of $N$ will define a twisted trace for $\aA$.  

In the following, we will often employ a very special {\it Grand Canonical} linear combination:
\begin{equation}
\langle O_1 \cdots O_k \rangle_\mu = \sum_{N=0}^\infty e^{\frac{2 \pi \mu N}{\epsilon_1}} \langle O_1 \cdots O_k  \rangle^{(N)} 
\end{equation}
which satisfies 
\begin{equation}
\partial_\mu \langle O_1 \cdots O_k \rangle_\mu = 2 \pi  \langle O_1 \cdots O_k  t_{0,0} \rangle_\mu
\end{equation}

\subsubsection{Higgs branch localization}
In the Higgs branch presentation, protected sphere correlation functions are computed by $N$-dimensional integrals over the eigenvalues of complexified holonomies:
\begin{equation}
\langle O_1 \cdots O_k \rangle^{(N)} = \frac{1}{N!} \left[\prod_{i=1}^N \int_{-\infty}^\infty d\sigma_i e^{2 \pi i \zeta \sigma_i}\right] \left[\prod_{i<j} 4 \sinh^2 \pi(\sigma_i - \sigma_j)\right] \langle O_1 \cdots O_k \rangle^{(N)}_{\mathrm{hyper}}
\end{equation}
where the ``free hypermultiplet'' correlation functions 
\begin{equation}
\langle O_1 \cdots O_k t_{n,m} \rangle^{(N)}_{\mathrm{hyper}}(\sigma_i)
\end{equation}
are computed by Wick contractions from Green functions
\begin{equation}
\langle X^a_b Y^c_d \rangle = \epsilon_1 \delta^a_d \delta^c_b \frac{1}{1+e^{2 \pi (\sigma_a- \sigma_c) }} \qquad \qquad \langle Y^c_d X^a_b \rangle = -\epsilon_1 \delta^a_d \delta^c_b \frac{1}{1+e^{2 \pi (\sigma_c- \sigma_a) }}
\end{equation}
and partition function
\begin{equation}
\langle 1 \rangle^{(N)}_{\mathrm{hyper}} = \prod_i \frac{1}{2 \cosh \pi \sigma_i} \prod_{i,j} \frac{1}{2 \cosh \pi(\sigma_i - \sigma_j)}
\end{equation}
The FI parameter $\zeta$ is given by 
\begin{equation}
\zeta = i \left(\frac12 + \frac{\epsilon_2}{\epsilon_1} \right)
\end{equation}
The integral remains convergent as long as 
\begin{equation}
-1<\mathrm{Re} \frac{\epsilon_2}{\epsilon_1} <0
\end{equation}
The finite $N$ partition function can be analytically continued outside of the strip, with poles along the real axis which become denser as $N$ increases. \footnote{The localization integral could also be modified by a 
mass $m$ for the adjoint hypermultiplet. This is equivalent, though, to the insertion of $e^{2 \pi m t_{1,1}}$ in correlation functions and does not add new information. The role of mass and FI parameters 
is exchanged in the mirror symmetric picture we employ later on in the Coulomb branch description of the algebra. }

The integral is straightforward but combinatorially daunting as a function of $N$. The systematic large $N$ expansion is poorly understood, but is expected to 
involve powers of $N^{-\frac12}$. Several calculations at the leading order in $N$ were done in  \cite{Mezei:2017kmw}.

\subsubsection{Grand canonical ensemble and free Fermi gas}
The large $N$ analysis is somewhat simpler in a grand-canonical ensemble, where one adds up correlation functions with different values of $N$:
\begin{equation}
\langle O_1 \cdots O_k \rangle_\mu = \sum_{N=0}^\infty e^{\frac{2 \pi \mu N}{\epsilon_1}} \langle O_1 \cdots O_k  \rangle^{(N)} 
\end{equation}
Then the large $N$ limit is probed at large positive values of $\frac{\mu}{\epsilon_1}$. 

The reason for the simplification is the Cauchy determinant identity, which allows one to combine the integration measure and the 
adjoint hypermultiplet partition function into a single determinant \cite{Kapustin:2010xq}:
\begin{equation}
\frac{\prod_{i<j} 4 \sinh^2 \pi(\sigma_i - \sigma_j)}{\prod_{i,j} 2 \cosh \pi(\sigma_i - \sigma_j)} = \sum_{s\in S_N} (-1)^s \prod_i \frac{1}{2 \cosh \pi (\sigma_i - \sigma_{s(i)})}
\end{equation}
and thus the correlation function as  
\begin{equation}
\langle O_1 \cdots O_k \rangle^{(N)} = \frac{1}{N!} \sum_{s\in S_N} (-1)^s  \left[\prod_{i=1}^N \int_{-\infty}^\infty \frac{ d\sigma_i e^{2 \pi i \zeta \sigma_i}}{4 \cosh \pi \sigma_i  \cosh \pi (\sigma_i - \sigma_{s(i)})}\right]  \frac{\langle O_1 \cdots O_k \rangle^{(N)}_{\mathrm{hyper}}}{\langle 1 \rangle^{(N)}_{\mathrm{hyper}}}
\end{equation}

The grand canonical partition function is then written as the partition function of a free Fermi gas \cite{Marino:2011eh} 
\begin{equation}
Z(\mu) \equiv \langle 1 \rangle_\mu = \det \left[1+ e^{\frac{2 \pi \mu}{\epsilon_1}} \hat \rho \right]
\end{equation}
with single-particle density operator $\hat \rho$ given by an integral operator with kernel 
\begin{equation}
\rho(\sigma, \sigma') =  \frac{e^{2 \pi i \zeta \sigma}}{4 \cosh \pi \sigma \cosh \pi (\sigma - \sigma')}
\end{equation}

The large $\frac{\mu}{\epsilon_1}$ limit of the partition function is well understood. We will review it momentarily. More general correlation functions also have a Fermi gas interpretation. Indeed, the grand canonical sum of expectation values of the form 
\begin{equation}
\frac{1}{N!} \sum_{s\in S_N} (-1)^s  \left[\prod_{i=1}^N \int_{-\infty}^\infty \frac{ d\sigma_i e^{2 \pi i \zeta \sigma_i}}{4 \cosh \pi \sigma_i  \cosh \pi (\sigma_i - \sigma_{s(i)})}\right]  \sum_{i_1<i_2 \cdots<i_n} f_n(\sigma_{i_a})
\end{equation}
can be written as the free Fermi gas expectation value of an operator acting on $n$ particles by multiplication by $f_n(\sigma_{i_a})$.

It is easy to see that a general correlation function with $n$ $X$ fields will insert in the integral a function of up to $n$ variables.

\subsubsection{Large $\mu$ limit}
The partition function has a very nice behaviour for large positive $\frac{\mu}{\epsilon_1}$ \cite{Nosaka:2015iiw}:
\begin{equation}
Z(\mu) \equiv \langle 1 \rangle_\mu  \sim Z_0(\epsilon_i) e^{\frac{4 \pi}{3 \sigma_3} \mu^3 + \frac{\pi \sigma_2}{4 \sigma_3} \mu}
\end{equation}
up to exponentially suppressed corrections. In particular, the perturbative expansion of the free energy truncates to a cubic polynomial in $\mu$, with no 
$\mu^{-1}$ corrections.

A striking feature of this perturbative expression is the triality invariance of the coefficients of $\mu^3$ and $\mu$. The whole partition function is definitely not 
triality invariant. Indeed, the original integral is invariant only under the trivial Weyl symmetry $\epsilon_2 \leftrightarrow \epsilon_3$.

It is also worth noticing that the prefactor $\frac{1}{\sigma_3}$ is the ``equivariant volume'' of the internal $\bC_{\epsilon_1} \times \bC_{\epsilon_2} \times \bC_{\epsilon_3}$
factor of the conjectural dual twisted M-theory background, and appears naturally as an overall prefactor in the twisted M-theory action. The parameter $\sigma_3$ thus plays a loop-counting 
role in twisted M-theory, and the perturbative expressions we find below are compatible with that interpretation. 

The leading coefficient $Z_0(\epsilon_i)$ has a conjectural expression
\begin{equation}
\log Z_0(\epsilon_i) = \frac12 A(1) + \frac14 A\left(- 2 \frac{\epsilon_2}{\epsilon_1} \right)+ \frac14 A\left(- 2 \frac{\epsilon_3}{\epsilon_1} \right)
\end{equation}
with 
\begin{equation}
A(z) = \frac{2 \zeta_3}{z \pi^2}\left(1-\frac{z^3}{16}\right)+ \frac{z^2}{\pi^2} \int_0^\infty \frac{x dx}{e^{z x}-1}\log (1-e^{- 2 x}) \qquad \qquad \mathrm{Re}\,z>0
\end{equation}
The range of definition of $A(z)$ covers the physical strip $-1<\mathrm{Re} \frac{\epsilon_2}{\epsilon_1} <0$. The function $A(z)$ is rather singular as $z$ approaches the imaginary axis,
so it is not clear that $Z_0(\epsilon_i)$ can be analytically continued beyond the physical strip. Within the physical strip, it is not triality invariant. In the following, we will 
typically strip $Z_0(\epsilon_i)$ off perturbative expressions by rescaling the correlation functions. 

Our main conjectural claim is that the grand-canonical correlation functions also have a truncated perturbative expansion at large positive $\frac{\mu}{\epsilon_1}$, i.e. the ratio 
\begin{equation}
Z(\mu)^{-1} \langle t_{m_1,n_1} \cdots t_{m_a,n_a} \rangle_\mu  
\end{equation}
approaches a polynomial in $\mu$ up to exponentially suppressed corrections. We can thus define a ``perturbative part'' of correlation functions:
\begin{equation}
\langle t_{m_1,n_1} \cdots t_{m_a,n_a} \rangle^{\mathrm{pert}}_\mu  \equiv e^{\frac{4 \pi}{3 \sigma_3} \mu^3 + \frac{\pi \sigma_2}{4 \sigma_3} \mu} \left[ Z(\mu)^{-1} \langle t_{m_1,n_1} \cdots t_{m_a,n_a} \rangle_\mu \right]_{\mathrm{pert}}
\end{equation}
Furthermore, because we have no inverse powers of $\mu$ in the expansion, we can simply set $\mu=0$ and encode the full $\mu$ dependence into $t_{0,0}$ insertions.

Experimentally, we find that the perturbative correlation functions $\langle t_{m_1,n_1} \cdots t_{m_a,n_a} \rangle^{\mathrm{pert}}_0$ are triality invariant. They are Laurent polynomials in  
$\sigma_3$ and polynomials in $\sigma_2$, of appropriate weight under the rescaling of $\epsilon_i$. They are natural candidates to match holographic calculations in some semiclassical 
saddle for twisted M-theory. 

In the remainder of this section, we will find a simple conjectural characterization of perturbative correlation functions. 

\subsubsection{Coulomb branch localization}
The Coulomb branch presentation of the $\aA$ algebra allows for an alternative localization calculation of the correlation functions. The calculation of general correlation functions is 
rather cumbersome, as it requires explicit ``Abelianized'' expressions for the monopole operators. Correlation functions of the commutative $d_n$ generators, though, 
are much simpler. We can write 

\begin{equation}
\langle d_{n_1} \cdots d_{n_k} \rangle^{(N)} =
\frac{1}{N!} \sum_{s\in S_N} (-1)^s  \left[\prod_{i=1}^N \int_{-\infty}^\infty \frac{ d\sigma_i}{4 \cosh \pi \sigma_i  \cosh \pi (\sigma_i - \sigma_{s(i)}+ \zeta)}\right] \prod_j \left[ \sum_i p_{n_j}(\sigma_i)\right]
\end{equation}

Here the $p_n(\sigma)$ polynomials are given by a generating series
\begin{equation}
\partial_z^2 \log \Gamma\left(\frac12 - i \sigma + z \right) = \sum_n \frac{p_n(\sigma)}{z^{n+1}}
\end{equation}

Because the $d_n$ generators commute, we can define a generating function
\begin{equation}
Z_d(\tau_i) = \langle e^{\sum_n \tau_n d_n} \rangle^{\mathrm{pert}}_0 \equiv e^{F_d(\tau_i)}
\end{equation}
where $F_d(\tau_i)$ is the generating function of connected correlation functions $\langle d_{n_1} \cdots d_{n_k}\rangle^{\mathrm{pert}}_c$.

With some numerical experimentation, we find a simple conjectural pattern: 
\begin{equation}
\langle d_{n_1} \cdots d_{n_k}\rangle^{\mathrm{pert}}_c = \sum_m c_{n_*;m} \sigma_2^m \sigma_3^{-\frac23 m + \frac13 \sum_i (n_i-1)}
\end{equation}
where the only non-vanishing terms have a power of $\sigma_3$ greater or equal $-1$, as expected for a loo-counting parameter. 

For example, we have
\begin{align}
\langle d_0 d_0 d_0\rangle^{\mathrm{pert}}_c &= \frac{1}{6 \pi^2 \sigma_3} \cr
\langle d_0 \rangle^{\mathrm{pert}}_c &= \frac{\pi \sigma_2}{4 \sigma_3} \cr
\langle d_2 d_0 d_0 d_0 d_0 \rangle^{\mathrm{pert}}_c &= -\frac{2}{\pi^4 \sigma_3} \cr
\langle d_2 d_0 d_0 \rangle^{\mathrm{pert}}_c &= -\frac{5 \sigma_2}{12 \pi^2 \sigma_3} \cr
\langle d_2 d_0 \rangle^{\mathrm{pert}}_c &= -\frac{1}{3 \pi^2} \cr
\langle d_2 \rangle^{\mathrm{pert}}_c &= -\frac{3 \sigma_2^2}{64 \sigma_3} \cr
\langle d_1 d_1 d_0 d_0 d_0 \rangle^{\mathrm{pert}}_c &= -\frac{2}{\pi^4 \sigma_3} \cr
\langle d_1 d_1 d_0 \rangle^{\mathrm{pert}}_c &= -\frac{\sigma_2}{3 \pi^2 \sigma_3} \cr
\langle d_1 d_1 \rangle^{\mathrm{pert}}_c &= -\frac{1}{12 \pi^2}-\frac{1}{64} 
\end{align}
etcetera.

\subsubsection{A recursion relation}
Inspection of the numerical data reveals a very simple recursion relation satisfied by $d_0$ insertions:
\begin{equation} \label{eq:rec}
\pi^2 \partial_{\tau_0}^2 F_d(\tau_i) + \left(\sum_n n \tau_n \partial_{\tau_{n-1}} \right)^2 F_d(\tau_i) = \sum \lambda_n \tau_n
\end{equation}
where $\lambda_n$ are functions of $\sigma_2$ and $\sigma_3$ only. This gives a quadratic relation on the perturbative correlation functions.
Experimentally, we find that this recursion relation combines with the twisted trace relations to uniquely fix all correlation functions!

In order to understand the origin of this recursion relation, it is useful to  consider the Fermi gas representation of the free energy
\begin{equation}
F_d(\tau_i)  = 2 \pi \Tr \log \left(1+ \hat \rho^{-1}_C [\tau_i] \right)
\end{equation}
where the Coulomb branch density operator is represented by the kernel 
\begin{equation}
\rho_C(\sigma, \sigma';\tau_i) =  \frac{e^{\sum_{n=0}^\infty \tau_n p_n (\sigma)} }{4 \cosh \pi \sigma \cosh \pi (\sigma - \sigma'+\zeta)} 
\end{equation}

In the following we will set $\epsilon_1$ to $1$ for simplicity. It can be restored by a trivial rescaling of $\sigma$ and $\tau_i$. 
We also assume large positive $\tau_0 \equiv 2 \pi \mu$. 

It is useful to observe that $\rho_C(\sigma, \sigma';\tau_i)$ has limited range, and if $|\sigma| \gg 1$ it is well approximated by 
\begin{equation}
\rho_C(\sigma, \sigma';\tau_i) \sim   \frac{e^{\pm \pi \sigma + \sum_{n=0}^\infty \tau_n p_n (\sigma)} }{2 \cosh \pi (\sigma - \sigma'+\zeta)} 
\end{equation}
up to exponential corrections.

The differential operator $i \sum_n n \tau_n \partial_{\tau_{n-1}}$ acts as a translation of the argument of the $p_n(\sigma)$ polynomials.
That means the combinations 
\begin{equation}
\pi \tau_0 \pm  i \sum_n n \tau_n \partial_{\tau_{n-1}}
\end{equation}
acts as a uniform translation on $\rho_C(\sigma, \sigma';\tau_i)$ in the regions $\pm \sigma \gg 1$. 

This suggests that the differential operator in the recursion relation \ref{eq:rec} annihilates the perturbative contribution to the free energy from the regions $\pi |\sigma| > \tau_0 - c$ where $c$ is some appropriate cutoff. It would be nice to complete this argument and show the origin of 
the linear source on the right hand side of \ref{eq:rec}.

We can give here some explicit examples of conjectural perturbative correlators. We have two-point functions 
\begin{align}
\langle t_{1}(u) t_{1}(v) \rangle^{\mathrm{pert}}_\mu &=\left[  \frac{1}{\sigma_3}\mu^2+ \frac{\sigma_2}{16 \sigma_3} \right] (u,v)\cr
\langle  t_{2}(u) t_{2}(v) \rangle^{\mathrm{pert}}_\mu &=\left[  \frac{16}{3 \pi \sigma_3}\mu^3+ \frac{4 \sigma_2}{3 \pi \sigma_3} \mu + \frac{1}{6 \pi^2} + \frac{1}{32} \right] (u,v)^2\cr
\langle t_{3}(u) t_{3}(v) \rangle^{\mathrm{pert}}_\mu &=\left[  \frac{3}{\sigma_3}\mu^4+ \frac{15 \sigma_2}{8 \sigma_3} \mu^2 + \frac{3}{2 \pi} \mu + \frac{27}{256} \frac{\sigma_2^2}{\sigma_3}\right] (u,v)^3
\end{align}
and three-point functions 
\begin{align}
\langle t_{1}(u) t_{1}(v) t_{2}(w) \rangle^{\mathrm{pert}}_\mu &=\left[  \frac{1}{\sigma_3}\mu^2+ \frac{\sigma_2}{16 \sigma_3} \right] (u,w)(v,w)\cr
\langle t_{1}(u) t_{2}(v) t_{3}(v) \rangle^{\mathrm{pert}}_\mu &=\left[  \frac{8}{\pi \sigma_3}\mu^3+ \frac{2 \sigma_2}{\pi \sigma_3} \mu + \frac{1}{4 \pi^2} + \frac{3}{64} \right] (u,w)(v,w)^2\cr
\langle t_{2}(u) t_{2}(v) t_{2}(v) \rangle^{\mathrm{pert}}_\mu&=\left[  \frac{32}{3\pi \sigma_3}\mu^3+ \frac{8 \sigma_2}{3\pi \sigma_3} \mu + \frac{1}{3 \pi^2} + \frac{1}{16} \right] (u,v)(u,w)(v,w)
\end{align}
We attach to the paper submission a Mathematica notebook which can compute general correlation functions. 

\section{M2 branes at an $A_1$ singularity}
The 3d ${\cal N}=4$ SQFT which flows to the world-volume theory of $N$ M2 branes at an $A_1$ singularity has two mirror descriptions.
The well known UV description as a stack of $N$ D2 branes in the presence of 2 D6 branes gives an  ADHM quiver with two flavours. A mirror description 
of the latter is a two-node necklace quiver with $U(N)$ gauge groups and a single flavour at the first node \cite{Porrati:1996xi, deBoer:1996mp}.

 We are interested in the Higgs branch protected correlators of the latter theory, or the 
Coulomb branch of the former. The corresponding algebra $\aA_N^{(2)}$ is conjecturally associated to twisted M-theory backgrounds where the $\bC \times \bC$ 
factor is replaced by the $A_1$ singularity or its deformation/resolution. 
\footnote{The opposite choices are also interesting, but are associated to a more intricate version of twisted M-theory, 
where the $A_1$ singularity lies in the $\Omega$ deformed directions. We will not study it here. }

The 3d ${\cal N}=4$ SCFT has an $SU(2)$ flavour symmetry inherited by the algebra $\aA_N^{(2)}$. It is the geometric isometry group of the $A_1$ singularity. In the Higgs branch description, 
it acts on the pair of bi-fundamental hypermultiplets. It is hidden in the Coulomb branch description, much as in the case of $\aA_N$.

If we denote the doublet of bifundamental hypermultiplets as $X_\alpha$, $Y_\alpha$ and the fundamentals as $I,J$, then the F-term relations take the schematic form 
\begin{align}
\epsilon^{\alpha \beta} X_\alpha Y_\beta + J I  &= z_1 1_{N \times N} \cr
\epsilon^{\alpha \beta} Y_\alpha X_\beta &= z_2 1_{N \times N}
\end{align}
In a manner similar to the case of $\aA_N$, we can reduce all operators to polynomials in the $SU(2)$ irreps of spin $k$:
\begin{equation}
\epsilon_1^{-1} \Tr X_{(\alpha_1} Y_{\alpha_2} \cdots X_{\alpha_{2k-1}} Y_{\alpha_{2k})}
\end{equation}
The $\Tr X_{(\alpha_1} Y_{\alpha_2)}$ are the $SU(2)$ generators. 
We will label the elementary operators $t^{(2)}_{a,b}$ by the $SU(2)$ quantum numbers as for $\aA_N$,
so that $\ell = \frac{a+b}{2}$ and $m = \frac{a-b}{2}$. Notice that $a-b$ is now always even. 

The Coulomb branch description $\aC^{(2)}_N$ of the algebra makes it easier to see its triality properties. 
Indeed, the Abelianized monopole operators have expressions which are simply identical to these of $\aC_N$, 
except that some operators are missing. More precisely, the elementary monopole operators in $\aC^{(2)}_N$
can be built within $\aC_N$ from the first few $d_n$ generators, together with $e_0$ and $f_1 + z f_0$. 
Conjecturally, these elements in $\aC$ generate the correct universal $\aC^{(2)}$.

It is easy to check that $e_0$, $f_1 + z f_0$, $d_1 + \frac{z}{2} d_0$ generate an $\mathfrak{su}(2)$ Lie algebra, 
which we identify with the global $SU(2)$ symmetry of $\aA_N^{(2)}$. Similarly, we embed 
\begin{equation}
t^{(2)}_{2n,0} =  t_{n,0} 
\end{equation}
and act with $t^{(2)}_{0,2}= f_1 + z f_0$ to build a full conjectural embedding of $\aA_N^{(2)}$ into $\aA_N$ and 
lift it to an embedding/definition of $\aA^{(2)}$ into $\aA$. \footnote{The embedding of $\aA_N^{(2)}$, or even $\aA_N^{(k)}$ below, into $\aA_N$ is somewhat unexpected from the ADHM definition of the algebras. From the $\Omega$-deformed M-theory perspective it can be motivated as encoding the relation between line defects placed at the origin of the $A_{k-1}$ singularity and line defects placed away from the origin. See \cite{Gaiotto:2019wcc} for a related discussion of algebra embeddings. It would be interesting to explore this point further.}

We can use that embedding to derive a concise conjectural presentation for the commutators defining $\aA^{(2)}$:
\begin{align}
[ t^{(2)}_{0,0}, t^{(2)}_{c,d} ] &= 0 \cr
[ t^{(2)}_{2,0}, t^{(2)}_{c,d} ] &= d\, t_{c+1,d-1}  \cr
[ t^{(2)}_{1,1}, t^{(2)}_{c,d} ] &= \frac12 (d-c) \,t_{c,d}  \cr
[ t^{(2)}_{0,2}, t^{(2)}_{c,d} ] &= - c \,t_{c-1,d+1}  \cr
[ t^{(2)}_{2 d,0}, t^{(2)}_{0,4} ] &= 4 d\, t^{(2)}_{2d-1,3} + \frac{2d(d-1)}{2d+1} (\sigma_2 d^2   -z^2)t^{(2)}_{2d-3,1}+\cr +& \sigma_3 \sum_{k=1}^{d-1}\frac{2 (d-k)(2k-1)(2d-k+1)}{2d+1} \left( t^{(2)}_{2k-2,0} t^{(2)}_{2d-2k-1,1}+t^{(2)}_{2d-2k-1,1}t^{(2)}_{2k-2,0}  \right)
\end{align}
Notice the explicit triality invariance. 
 
The algebra $\aA_N^{(2)}$ is a deformation of the algebra of Hamiltonian symplectomorphisms on $\frac{\bC \times \bC}{\bZ_2}$:
\begin{equation}
[ t^{(2)}_{a,b}, t^{(2)}_{c,d} ] = \frac12 (ad-bc)t^{(2)}_{a+c-1,b+d-1} + O(\epsilon_i)
\end{equation} 

\subsection{Correlation functions}
We can solve the twisted trace conditions in the same manner as for the  case of $\aA$, conjecturally reducing any correlation function to a linear combination of 
$\langle (t^{(2)}_{2,0})^{\sum_i n_i} \prod_i t_{0,2n_i} \rangle$ extremal correlators.

The localization expressions for the correlation functions can be also manipulated in a familiar way. On the Higgs branch side, 
we can employ the Cauchy identity
\begin{equation}
\frac{\prod_{i<j} 4 \sinh \pi(\sigma_i - \sigma_j)\sinh \pi(\sigma'_i - \sigma'_j)}{\prod_{i,j} 2 \cosh \pi(\sigma_i - \sigma'_j)} = \sum_{s\in S_N} (-1)^s \prod_i \frac{1}{2 \cosh \pi (\sigma_i - \sigma'_{s(i)})}
\end{equation}
to arrive to a standard Fermi gas description of the grand canonical partition function, involving the integral operator with kernel
\begin{equation}
\rho^{(2)}_H(\sigma, \sigma'') =  \frac{e^{2 \pi i \zeta_1 \sigma}}{2 \cosh \pi \sigma} \int_{-\infty}^\infty  \frac{e^{2 \pi i \zeta_2 \sigma'}}{4\cosh \pi (\sigma - \sigma')\cosh \pi (\sigma' - \sigma'')}
\end{equation}
where the parameters $\zeta_1$, $\zeta_2$ are (affine) linearly related to $z_1$, $z_2$ or $\epsilon_2$, $z$. 

On the Coulomb branch side, one has a Fermi gas description with Fourier-transformed kernel: 
\begin{equation}
\rho^{(2)}_C(\sigma, \sigma') = \frac{1}{8 \cosh \pi \sigma \cosh \pi (\sigma +\zeta')\cosh \pi (\sigma - \sigma'+\zeta)}
\end{equation}
where $\zeta' = -i z$. The $d_n$ insertions are controlled by the same $p_n(\sigma)$ polynomials. 

We define the grand canonical perturbative correlation functions as before. The main difference is that now we have 
\begin{equation}
Z^{(2)}(\mu) \sim Z^{(2)}_0(\epsilon_i) e^{\frac{2 \pi}{3 \sigma_3} \mu^3 + \frac{\pi z^2}{2 \sigma_3} \mu}
\end{equation}

Experimentally, we find that the perturbative, grand canonical perturbative connected correlators of the $d_n$ generators satisfy a recursion relation 
\begin{equation} \label{eq:rec2}
4 \pi^2 \partial_{\tau_0}^2 F_d(\tau_i) + \left(\sum_n n \tau_n \partial_{\tau_{n-1}} \right)^2 F_d(\tau_i) = \sum \lambda_n \tau_n
\end{equation}
analogous to \ref{eq:rec}, which determines them uniquely. 

We also find a simple recursion for the $z$ dependence: 
\begin{equation} \label{eq:rec2z}
\partial_{z} F_d(\tau_i) + \frac12 \left(\sum_n n \tau_n \partial_{\tau_{n-1}} \right) F_d(\tau_i) = \sum \lambda'_n \tau_n.
\end{equation}

\section{M2 branes at an $A_k$ singularity}
The SCFT associated to $M2$ branes at an $A_k$ singularity can be obtained either from a necklace quiver of $k+1$ nodes with a single flavour or as 
an ADHM quiver with $k+1$ flavours. We consider the Higgs branch correlators in the former theory, or Coulomb branch correlators in the latter
and take the uniform-in-$N$ limit. 

We do not have a concise presentation of the resulting algebra. We expect it to admit generators $t^{(k)}_{a,b}$ with $a-b$ multiple of $k$, as well as a triality invariant presentation which deforms 
\begin{equation}
[ t^{(k)}_{a,b}, t^{(k)}_{c,d} ] = \frac12 (ad-bc)t^{(k)}_{a+c-1,b+d-1} + O(\epsilon_i)
\end{equation} 
depending on $\sigma_2$, $\sigma_3$ and the $k$ deformation parameters $z_i$. Using the Coulomb branch description, one can conjecturally embed into 
the Coulomb branch for the theory with no flavours \cite{Gaiotto:2019wcc}. The embedding includes $e_0$, the first few $d_n$'s and 
\begin{equation}
f_k + \left[ \sum_i z_i \right] f_{k-1} + \left[ \sum_{i<j} z_i z_j \right] f_{k-2} + \cdots+ \left[ \prod_i z_i \right] f_0
\end{equation}

Coulomb branch correlators of the $d_n$'s can be computed as before from localization integrals and the Fermi Gas construction. 
We expect recursion relations of the form
\begin{equation} \label{eq:reck}
k^2 \pi^2 \partial_{\tau_0}^2 F_d(\tau_i) + \left(\sum_n n \tau_n \partial_{\tau_{n-1}} \right)^2 F_d(\tau_i) = \sum \lambda_n \tau_n
\end{equation}
as well as 
\begin{equation} \label{eq:reckz}
k \partial_{z_a} F_d(\tau_i) + \left(\sum_n n \tau_n \partial_{\tau_{n-1}} \right) F_d(\tau_i) = \sum \lambda'_{a,n} \tau_n.
\end{equation}

Given an explicit presentation of the algebra, one should be able to compute all correlation functions via twisted trace relations and 
the recursion relations. We leave it for future work.

\section{Hidden triality in the Schur index}
The final collection of protected correlation functions we will consider will be the Schur indices for line defect junctions in 4d ${\cal N}=4$ SYM
with $U(N)$ gauge group.
 
Recall that the Schur index is a specialization of the superconformal index which is available for any 4d ${\cal N}=2$ SQFT \cite{Gadde:2011uv}. 
It can be thought of as a supersymmetric partition function on $S^1 \times S^3$. It can be decorated by collections of BPS line defects 
wrapping the $S^1$ factor of the $S^1 \times S^3$ space-time geometry \cite{Dimofte:2011py}. From the point of view of the superconformal index, 
the resulting correlation functions count local operators at supersymmetric junctions of half-BPS line defect. 

These ``Schur correlation functions'' \footnote{Not to be confused with a different, and presumably incompatible, ``Higgs branch'' generalization of the Schur index,
which inserts local operators rather than line defects and gives rise to torus conformal blocks of a certain chiral algebra \cite{Dedushenko:2019yiw}. These also are potential targets for twisted holography calculations \cite{Costello:2018zrm}, but will be discussed elsewhere.} have many properties in common with Coulomb branch sphere correlation functions in 3d ${\cal N}=4$ SQFTs. In particular, the OPE of line defects gives a quantization of the algebra of functions on the Coulomb branch of the 4d theory compactified on a circle. The Schur correlation functions behave as a twisted trace on the algebra, with a twist which is trivial for 4d  SCFTs.  

From this point on, with ``Coulomb branch'' we will always refer to the Coulomb branch of the 4d theory compactified on the circle, and with 
``quantum Coulomb branch algebra'' we will always refer to the non-commutative algebra of line operators which arise from a twisted circle compactification 
on the circle \cite{Gaiotto:2010be}, which controls the OPE in the Schur correlators. 

We are interested in the Schur correlation functions of 4d ${\cal N}=4$ $U(N)$ SYM, possibly deformed by an ${\cal N}=2^*$ flavour fugacity. In order to relate this to the M-theory considerations in the previous sections, we may notice a few facts: 
\begin{itemize}
\item The Coulomb branch of ${\cal N}=2^*$ $U(N)$ SYM \cite{Donagi:1995cf} (in the generic complex structure relevant here) is a multiplicative analogue of the Higgs or Coulomb branch algebras of the M2 brane theory. The $X$ and $Y$ adjoint matrices are replaced by $GL(N)$ group elements $U$ and $V$ and the moment map relation is replaced by the constraint 
\begin{equation}
\zeta U V - \zeta^{-1} V U = J I
\end{equation}  
\item The analogy with the M2 brane theory becomes stronger after a standard string duality, mapping D3 branes wrapping a circle to M2 branes with a transverse $\bC^* \times \bC^*$ geometry. Wilson loops map to BPS operators charged under rotations of one $\bC^*$ factor. 't Hooft loops map to 
BPS operators charged under rotations of the second $\bC^*$ factor. S-duality acts geometrically on $\bC^* \times \bC^*$ as $u \to u^a v^b$, $v \to u^c v^d$. 
\item If we take the uniform-in $N$ limit and turn off the mass deformation parameters, the Poisson algebra of functions on the 
Coulomb branch becomes the universal enveloping algebra $U(\mathfrak{t})$ of the Lie algebra of hamiltonian symplectomorphisms of $\bC^* \times \bC^*$.
The uniform-in-$N$ limit of the quantum, mass deformed Coulomb branch algebra is a two-parameter deformation of that. It is a natural candidate for the Koszul dual to the algebra of observables of twisted M-theory on $\bR \times \bC^* \times \bC^*$. 
\end{itemize}

We will denote as $L_{0,1}$ the BPS operator associated to the fundamental Wilson loop, $L_{0,-1}$ the anti-fundamental one and as 
$L_{m,n}$ their S-duality images, with $m$,$n$ co-prime.

We plan to make manifest a large $N$ hidden triality of both OPE and correlation functions, mixing the quantization parameter $q$ and the complexified fugacity $\zeta$ for the ${\cal N}=2^*$ deformation.

\subsection{The quantum Coulomb branch algebra} 
The quantum Coulomb branch algebra $\aB_N$ can be presented in an Abelianized form \cite{Drukker:2009id,Alday:2009fs,Gomis:2011pf,Bullimore:2015lsa}, where the Wilson line defects are given as
symmetric polynomials in gauge fugacities $\sigma_i$, such as the fundamental and anti-fundamental
\begin{equation}
L_{0,1}= \sum_i \sigma_i \qquad \qquad  L_{0,-1}= \sum_i \sigma_i^{-1}
\end{equation}
More general Wilson-'t Hooft operators are given as intricate difference operators acting on the $\sigma_i$ by linear combinations of 
$\sigma_i \to q^n \sigma_i$ transformations. Explicit expressions are available for the elementary 't Hooft operators $L_{\pm 1,n}$ of magnetic charge $\pm 1$ and general electric charge $n$ (aligned to the magnetic charge) as Macdonald difference operators. 

It is possible to find explicit transformations manifesting the $SL(2,\bZ)$ S-duality symmetry of $\aB_N$. For example, one could realize the transformation kernels for the $S$ transformation as 
supersymmetric indices \cite{Gang:2013sqa,Cordova:2016uwk} of the 3d ${\cal N}=4$ $T[U(N)]$ gauge theories \cite{Gaiotto:2008ak}, generalizing the classical results of \cite{Gaiotto:2013bwa}. 
Appropriate S-duality transformations map the $L_{0,\pm 1}$ operators into the $L_{\pm 1,n}$. Alternative, manifestly S-dual presentations of the algebra in terms of skeins on a punctured torus are 
also available \cite{Bullimore:2013xsa}.

Mathematically, the algebra $\aB_N$ should coincide with the spherical DAHA  algebra $\text{\bf S\"H}_N$ and the uniform-in-$N$ limit is presented 
in an explicitly $SL(2,\bZ)$-invariant and triality invariant form as $\text{\bf S\"H}_\infty$ in reference \cite{2009arXiv0905.2555S}.
Following that reference, will normalize 
\begin{equation}
\ell_{0,\pm1} = \frac{1}{q^{-\frac12}- q^{\frac12}} L_{0,\pm 1} \qquad \qquad \ell_{\pm1,n} = \frac{1}{q^{-\frac12}- q^{\frac12}} L_{\pm 1,n}
\end{equation}
This rescaling is compatible with S-duality. It is analogous to the $\epsilon_1^{-1}$ factor in the definition of $t_{n,m}$ for the M2 brane algebra. 

It is instructive to rediscover some of the relations from \cite{2009arXiv0905.2555S}.
From the definition, we find 
\begin{equation}
[\ell_{1,n},\ell_{0,\pm 1}] = \pm \ell_{1,n\pm1} \qquad \qquad [\ell_{-1,n},\ell_{0,\pm 1}] = \mp \ell_{-1,n\pm1} 
\end{equation}

Because of S-duality, it must be possible to define $\ell_{m,n}$, with $m$ and $n$ coprime, such that if $m n' -n m'=1$ we have 
\begin{equation}
[\ell_{m,n},\ell_{m',n'}] =  \ell_{m+m',n+n'}  
\end{equation}
and furthermore $[\ell_{m,n},\ell_{-m,-n}]=0$. Such $\ell_{m,n}$ can be found explicitly by applying the above relation recursively, starting from 
the expressions for $\ell_{0,\pm 1}$ and $\ell_{\pm 1,n}$. We denote $\ell_{m,n}$ with $m$ and $n$ coprime as ``minimal'' generators.

Using these definitions, we can then compute more general commutators, such as
\begin{equation}
[\ell_{1,0},\ell_{1,3}]= (q_1 + q_2 + q_3) \ell_{2,3} + (1-q_1)(1-q_2)(1-q_3) \ell_{1,1}\ell_{1,2}
\end{equation}
where we defined $q_1 = q$, $q_2 = \zeta$, $q_3 = q_1^{-1} q_2^{-1}$. We can also write that as 
\begin{equation}
[\ell_{1,0},\ell_{1,3}]=  (q_1^{-1} + q_2^{-1} + q_3^{-1}) \ell_{1,1}\ell_{1,2}-(q_1 + q_2 + q_3) \ell_{1,2}\ell_{1,1}
\end{equation}
which implies the S-dual image
\begin{equation}
[\ell_{a,b},\ell_{a+3 c,b+3d}]=  (q_1^{-1} + q_2^{-1} + q_3^{-1}) \ell_{a+c,b+d}\ell_{a+2 c,b+2 d}-(q_1 + q_2 + q_3) \ell_{a+2 c,b+2 d}\ell_{a+c,b+d}
\end{equation}
whenever $(a d - b c) = \pm 1$. These commutators are invariant under triality transformations permuting the $q_i$, as expected. 

Another important observation is that commutators $[\ell_{1,n},\ell_{-1,n'}]$ give generators $\ell_{0,n+n'}$ built from Wilson line defects of higher charge, which all commute with $\ell_{0,\pm 1}$. With the help of S-duality, we can get canonical definitions of $\ell_{n,m}$ for non-coprime $n$,$m$. 

When $\zeta=1$, it is known that the quantum Coulomb branch algebra reduces to the symmetric product of $N$ copies of the quantum torus algebra 
$x y = q y x$. Correspondingly, $\aB$ reduces to the universal enveloping algebra of the Lie algebra 
\begin{equation}
[\ell_{m,n},\ell_{m',n'}] = [m n' - n m']_q \ell_{m+m',n+n'}
\end{equation}
of the quantum torus algebra. Setting $q\to 1$ as well gives the universal enveloping algebra $U(\mathfrak{t})$ of the Lie algebra $\mathfrak{t}$ of Hamiltonian symplectomorphisms of $\bC^* \times \bC^*$:
\begin{equation}
[\ell_{m,n},\ell_{m',n'}] = (m n' - n m') \ell_{m+m',n+n'}
\end{equation}
as desired. 

Next, we will test the triality properties of the correlators. We can begin by studying somewhat heuristically the consequences of the trace relations. 
\subsection{Reduction to Wilson line correlators}
We have not worked out the precise space of solutions of trace relations. We expect the analysis to proceed in a manner analogous as for 
$U(\mathfrak{t})$. 

We can give an example of such a reduction $U(\mathfrak{t})$. We have relations such as  
\begin{equation}
\langle \ell_{0,1} \ell_{0,1} \ell_{0,-2} \rangle = \frac12 \langle \ell_{0,1} \ell_{0,1} [\ell_{-1,-1},\ell_{1,-1}] \rangle = \frac12 \langle \ell_{-1,0} \ell_{0,1} \ell_{1,-1}\rangle+\frac12 \langle \ell_{0,1}  \ell_{-1,0} \ell_{1,-1}\rangle
\end{equation}
which allows us to write 
\begin{equation}
\langle \ell_{0,1}  \ell_{-1,0} \ell_{1,-1}\rangle=\langle \ell_{0,1} \ell_{0,1} \ell_{0,-2} \rangle  + \langle \ell_{-1,1}\ell_{1,-1}\rangle
\end{equation}
Because of S-duality, 
\begin{equation}
\langle \ell_{-1,1}\ell_{1,-1}\rangle = \langle \ell_{0,1}\ell_{0,-1}\rangle
\end{equation}
and thus we have reduced the non-trivial three-point function $\langle \ell_{0,1}  \ell_{-1,0} \ell_{1,-1}\rangle$ to a linear combination  of Wilson line correlation functions. 

It is reasonable to hope that all correlation functions may be expressible as linear combinations  of Wilson line correlation functions, perhaps satisfying some further constraints. This would be analogous to the reduction to correlation functions of the $d_n$ operators in the 3d case. 

We thus focus on Schur correlation functions of Wilson lines. 
\subsection{Wilson line correlation functions}
In the presence of Wilson line defect insertions, the Schur index is a contour integral of a ration of theta functions multiplied by appropriate characters of the gauge group:
\begin{equation}
\langle \prod_a W_{R_a} \rangle^{(N)} = \frac{1}{N!}  \left[\prod_{i=1}^N \oint_{|\sigma_i|=1} ds_i\right] \frac{(q)_\infty^{3N}(\tau) \prod_{i<j} \theta(\sigma_i/\sigma_j;q)\theta(\sigma_j/\sigma_i;q)}{\prod_{i,j} \theta(\sigma_i/\sigma_j \zeta^{-1};q)} \prod_a \chi_{R_a}(\sigma_*)
\end{equation}
with $|q|<|\zeta|^{-1}<1$ and 
\begin{equation}
\theta(\zeta;q) = (\zeta^{\frac12}-\zeta^{-\frac12})\prod_{n=1}^\infty (1-q^n)(1-\zeta q^n)(1-\zeta^{-1} q^n) = \sum_{n \in \bZ} (-1)^n \zeta^{n+\frac12} q^{\frac12(n^2 + n)}
\end{equation}

We have 
\begin{equation}
\theta(e^{ 2 \pi i n} q^m \zeta;q) = (-1)^{n+m} q^{-\frac{m^2}{2}} \zeta^{-m}\theta(\zeta;q)
\end{equation}
and $\theta(\zeta^{-1};q) = - \theta(\zeta;q)$. 

In order to proceed, we would like some analogue of a grand canonical partition function. 
Consider the following function:
\begin{equation}
G(\zeta,u;q) = \frac{\theta(\zeta u;q) (q)_\infty^3}{\theta(\zeta;q) \theta(u;q)}
\end{equation}
It satisfies
\begin{align}
G(e^{ 2 \pi i n} q^m \zeta,u;q) &= u^{-m} G(\zeta,u;q) \cr
G(\zeta,e^{ 2 \pi i n} q^m u;q) &= \zeta^{-m} G(\zeta,u;q)
\end{align}

It has a useful Fourier expansion valid in the fundamental region $|q|<|\zeta|<1$:
\begin{equation}
G(\zeta,u;q) =- \sum_n \frac{\zeta^n}{1-u q^n}
\end{equation}
Notice 
\begin{equation}
G(q \zeta^{-1},u;q) =-u^{-1} G(\zeta,u^{-1};q)
\end{equation}

Among other things, the $G(\zeta,u;q)$ function is used to define the two point function of a complex fermion on the torus
coupled to a Spin$_c$ bundle. Because of bosonization, it obeys an interesting Frobenius determinant formula
\begin{equation}
\det_{i,j}  G(v_i/w_j,u;q) = \frac{\theta(u \prod_i v_i/w_i;q)}{\theta(u;q)}\frac{(q)_\infty^{3N} \prod_{i<j} \theta(v_i/v_j;q)\theta(w_j/w_i;q)}{\prod_{i,j} \theta(v_i/w_j;q)}
\end{equation}
We are particularly interested in the case where $w_i = v_i \zeta$, in which case we have 
\begin{equation}
\det_{i,j}  G(v_i/v_j\zeta^{-1},u;q) = \frac{\theta(u \zeta^{-N};q)}{\theta(u;q)}\frac{(\eta)q)_\infty^{3N}\prod_{i<j} \theta(v_i /v_j;q)\theta(v_j /v_i;q)}{\prod_{i,j} \theta(v_i/v_j\zeta^{-1};q)}
\end{equation}
so that \cite{Bourdier:2015wda}
\begin{equation}
\frac{\theta(u \zeta^N;q)}{\theta(u;q)}\langle \prod_a W_{R_a} \rangle^{(N)} = \frac{1}{N!}  \left[\prod_{i=1}^N \oint_{|\sigma_i|=1} ds_i\right] \det_{i,j}  G(\sigma_i/ \sigma_j \zeta^{-1},u;\tau) \prod_a \chi_{R_a}(\sigma_*)
\end{equation}

This means we can define grand canonical correlation functions as 
\begin{equation}
\langle \prod_a W_{R_a} \rangle_{\xi,u}  = \sum_N \xi^N \frac{\theta(u \zeta^{-N};q)}{\theta(u;q)} \langle \prod_a W_{R_a} \rangle^{(N)} 
\end{equation}
and they will have a free Fermi gas interpretation. Notice that we introduced two new fugacities, $\xi$ and $u$. This is a bit redundant, 
but will be very useful. 

\subsection{Explicit examples and triality invariance}
The single particle density operator $\hat \rho$ is an integral operator which acts on functions on $S^1$ as convolution with $G(\zeta,u;q)$. 
In Fourier transform, it acts on functions on $\bZ$ as multiplication by $-\frac{\zeta^{-n}}{1-u q^n}$. 

As a consequence, we can immediately compute the grand canonical partition function
\begin{equation}
Z(\xi,u;\zeta,q) = \prod_{n=-\infty}^\infty \left(1- \frac{\xi \zeta^{-n}}{1-u q^n} \right) =\prod_{n=-\infty}^\infty \frac{1-u q^n- \xi \zeta^{-n} }{1-u q^n}
\end{equation}
in the fundamental region $|q|<|\zeta|^{-1}<1$.

Notice that the naive Weyl symmetry acting on the ${\cal N}=2^*$ fugacity $\zeta \to q^{-1} \zeta^{-1}$, i.e. $q_2 \leftrightarrow q_3$, must be accompanied by a 
redefinition of the auxiliary fugacities:
\begin{equation}
Z(\xi,u;q^{-1} \zeta^{-1},q) = \prod_{n=-\infty}^\infty \frac{1-u q^n- \xi \zeta^{n} q^n}{1-u q^n}= \prod_{n=-\infty}^\infty \frac{1-u^{-1} q^{n}+ \xi u^{-1} \zeta^{-n} }{1-u^{-1} q^{n}}
\end{equation}
i.e. $Z(\xi,u;\zeta,q) = Z(-\xi u^{-1} ,u^{-1};q^{-1} \zeta^{-1},q)$.

We can also compute some correlation functions \cite{Drukker:2015spa}. The Wilson line operators map to very simple operators in the Fermi gas description. 
For example, $W_{0,\pm 1} = \sum_i \sigma_i^{\pm 1}$ maps to an operator acting on single fermions. In Fourier transform, the 
fermion modes are labelled by an integer, and  $W_{0,\pm 1}$ acts on the integer label as a shift by $\pm 1$.

We find 
\begin{equation}
Z(\xi,u;\zeta,q)^{-1} \langle W_{0,1} W_{0,-1} \rangle = \mathrm{Tr} \frac{\xi \hat \rho}{1+ \xi \hat \rho} \hat W_{0,1} \hat W_{0,-1} -  \mathrm{Tr} \frac{\xi \hat \rho}{1+ \xi \hat \rho} \hat W_{0,1}\frac{\xi \hat \rho}{1+ \xi \hat \rho} \hat W_{0,-1}
\end{equation}
where $\hat W_{0,\pm 1}$ acts by shift $f_n \to f_{n\pm 1}$. As a consequence, we have 
\begin{equation}
q^{-1} (1-q)^2 \frac{\langle \ell_{0,1} \ell_{0,-1} \rangle}{Z(\xi,u;\zeta,q)} = - \sum_n \frac{\xi \zeta^{-n}}{1-u q^n-\xi \zeta^{-n}} - \sum_n \frac{\xi^2 \zeta^{-2n-1}}{\left(1-u q^n-\xi \zeta^{-n}\right)\left(1-u q^{n+1}-\xi \zeta^{-n-1}\right)}
\end{equation}
which is also invariant under the Weyl symmetry $\zeta \to q^{-1} \zeta^{-1}$, $u \to u^{-1}$, $\xi \to -\xi u^{-1}$. 

We can manipulate that expression in two ways, as
\begin{equation}
q^{-1} (1-q)^2 Z(\xi,u;\zeta,q)^{-1} \langle \ell_{0,1} \ell_{0,-1} \rangle = - \sum_n \frac{\xi \zeta^n\left(1-u q^{n+1}\right) }{\left(1-u q^n-\xi \zeta^n\right)\left(1-u q^{n+1}-\xi \zeta^{n+1}\right)} 
\end{equation}
or 
\begin{equation}
q^{-1} (1-q)^2 Z(\xi,u;\zeta,q)^{-1} \langle t_{0,1} t_{0,-1} \rangle = - \sum_n \frac{\xi \zeta^{n+1}\left(1-u q^{n}\right) }{\left(1-u q^n-\xi \zeta^n\right)\left(1-u q^{n+1}-\xi \zeta^{n+1}\right)} 
\end{equation}
and then take a linear combination 
\begin{equation}
q^{-1} (1-q)^2(\zeta-1) Z(\xi,u;\zeta,q)^{-1} \langle t_{0,1} t_{0,-1} \rangle = (q-1) \sum_n \frac{u q^{n}\xi \zeta^{n+1} }{\left(1-u q^n-\xi \zeta^n\right)\left(1-u q^{n+1}-\xi \zeta^{n+1}\right)} 
\end{equation}
to a neat final form
\begin{equation}
Z(\xi,u;\zeta,q)^{-1}  \langle t_{0,1} t_{0,-1} \rangle =- \frac{1}{(1-\zeta)(1-q)} \sum_n \frac{u q^{n+1}\xi \zeta^{-n} }{\left(1-u q^n-\xi \zeta^{-n}\right)\left(1-u q^{n+1}-\xi \zeta^{-n-1}\right)} 
\end{equation}
Here we get to the crucial point: this expression converges for most values of $\zeta$, $q$, except at $|\zeta||q|=1$. It can be thought of as an analytic continuation of the original expression. It has a manifest non-trivial triality symmetry $q \leftrightarrow \zeta$, $\xi \leftrightarrow \mu$, which together with the $\zeta \to q^{-1} \zeta^{-1}$, $u \to u^{-1}$, $\xi \to -\xi u^{-1}$ Weyl transformation generates a full $S_3$ triality group. 

We can also take a different linear combination 
\begin{equation}
Z(\xi,u;\zeta,q)^{-1}  \langle t_{0,1} t_{0,-1} \rangle = \frac{1}{(1-q)(1-q^{-1} \zeta^{-1})} \sum_n \frac{\xi \zeta^{-n-1}}{\left(1-\mu q^n-\xi \zeta^{-n}\right)\left(1-\mu q^{n+1}-\xi \zeta^{-n-1}\right)} 
\end{equation}
which converges away from $|\zeta|=1$.

In conclusion, the correlation function is well-defined away from $|\zeta|=|q|=1$ and triality invariant!

Parsing through the definitions of the $\ell_{0,n}$, we find that the expression 
\begin{equation}
\tilde \ell_{0,2} = \frac{1}{q^{-1}-q} \sum_i \sigma_i^2
\end{equation}
is a triality-invariant linear combination of $\ell_{0,2}$ and $\ell_{0,1}^2$. 

We have again a nice and triality-invariant expression
\begin{equation}
Z(\xi,u;\zeta,q)^{-1}  \langle \tilde \ell_{0,2} \tilde \ell_{0,-2} \rangle = -\frac{1}{(1-\zeta^2)(1-q^2)} \sum_n \frac{\mu q^{n+2}\xi \zeta^{-n} }{\left(1-\mu q^n-\xi \zeta^{-n}\right)\left(1-\mu q^{n+2}-\xi \zeta^{-n-2}\right)} 
\end{equation}

We can also compute with some work 
\begin{align}
 \frac{ \langle \ell_{0,1} \ell_{0,1} \tilde \ell_{0,-2} \rangle}{Z(\xi,u;\zeta,q)} = \sum_n \frac{(1-\zeta)^{-1}(1-q)^{-1}(1-q\zeta)^{-1} \mu q^{n+2}\xi \zeta^{-n} }{\left(1-\mu q^n-\xi \zeta^{-n}\right)\left(1-\mu q^{n+1}-\xi \zeta^{-n-1}\right)\left(1-\mu q^{n+2}-\xi \zeta^{-n-2}\right)} 
\end{align}
which is again triality invariant.

Based on these examples. it is natural to conjecture that the normalized grand canonical Schur correlators are all triality invariant.

\noindent {\bf Acknowledgements.} We thank Tadashi Okazaki for collaboration at an early stage of the project. 
We thank Jihwan Oh, Yehao Zhou for providing proofs of some conjectural commutation relations.  
This research was supported in part by a grant from the Krembil Foundation. J.A. and D.G. are supported by the NSERC
Discovery Grant program and by the Perimeter Institute for Theoretical
Physics. Research at Perimeter Institute is supported in part by the Government of Canada through the Department of Innovation, Science and Economic Development Canada and by the Province of Ontario through the Ministry of Colleges and Universities.

\appendix
\section{Numerics}

We have conducted extensive numerical tests of the conjectured properties of the grand-canonical correlation functions, which we summarize in this appendix.  These tests were performed by considering the ``Fermi gas'' expressions for the the Coulomb branch presentation of the BPS algebra explained above.  The relevant functional operators were discretized and represented as matrices, and the relevant functional operator traces were calculated numerically using \textsc{Mathematica}.  For example, the density matrix, $\rho$ acts on elements of the Hilbert space of functions as a convolution.  

To represent $\rho$ numerically, we note first that the kernel of this convolution is small far away from the diagonal.  Because of this, we can safely cutoff the region of integration for the convolution at some ``large'' $L>0$, and integrate only between $-L$ and $L$. What constitutes ``large'' grows linearly with $\mu$. For the correlation functions and range of $\mu$ we considered here, $L\sim 12$ appears to works well.  

The integration is then discretized, with the range $[-L,L]$ being represented by a large number $M$, of sample points spaced evenly within the interval. The ratio $M/L$ has to be sufficiently large to cover the the spread of the kernel in momentum space, which also grows linearly in $\mu$. For the range we consider below, we found that $M \sim 1000$ works well.  Similar discretizations are done for the other relevant operators (e.g. the ones representing the various $d_i$'s).  Traces of combinations of these matrices can then be computed numerically. 

In this way, data was generated for several of the 1 and 2 point correlation functions of $d_i$ operators for small $i$.  These numerical results were then examined to verify qualitative features (e.g. triality invariance at large $\mu$, the absence of $\mu^{-1}$ contributions), and the perturbative part of the results was compared with the expressions obtained from the conjectured recursion relations.

As an illustration of some of the checks we have done, the accuracy of the fit between the conjectured expressions for the perturbative part of $\langle d_2 \rangle$ and the numerically obtained data is exhibited in Figure 1.  Similar comparisons for $\langle d_2 d_4\rangle$ are shown in Figure 2.  In both cases, at small $\mu$, there is an exponentially decaying mismatch between conjecture and data.  This is due to the presence of nonperturbative corrections to the grand-canonical correlation functions.  Importantly, we find that, with the inclusion of an exponentially decaying term to account for the expected nonperturbative effects, there are no $\mu^{-1}$ contributions.  At $\mu \approx 1$, we have very good agreement, since the nonperturbative corrections are negligible there.  At larger values of $\mu$, the difference grows again.  However, this does not represent anything physical.  Rather, it is due to the error in the numerical calculation introduced by finite $L$ and $M/L$. Indeed, we can increase or reduce this part of the error by adjusting the resolution with which the relevant functional operators are represented numerically.

\begin{figure}
\centering
\begin{subfigure}{0.5\textwidth}
\centering
\includegraphics[width=0.9\textwidth]{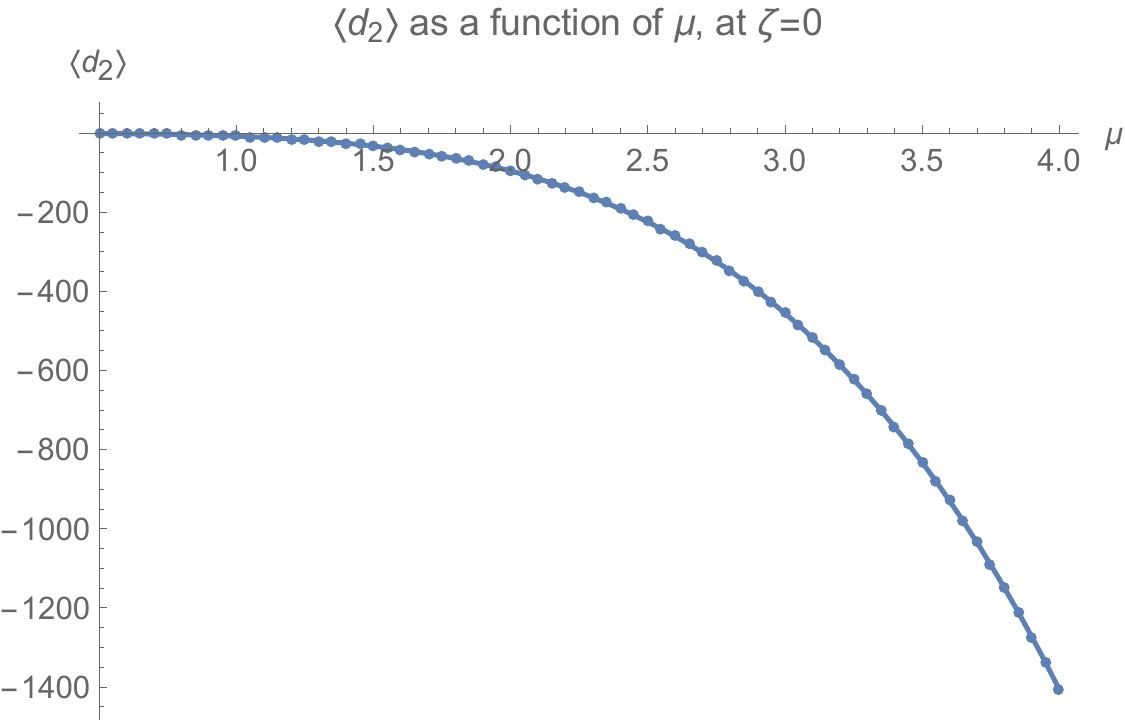}
\end{subfigure}%
\begin{subfigure}{0.5\textwidth}
\centering
\includegraphics[width=0.9\textwidth]{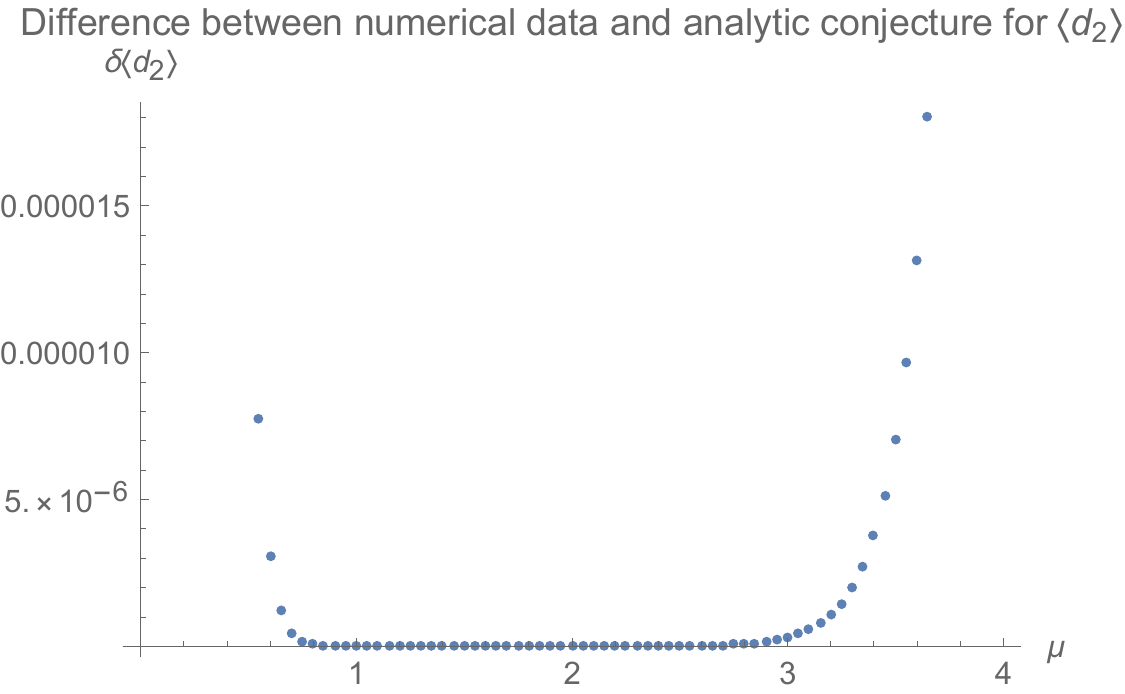}
\end{subfigure}
\begin{subfigure}{0.5\textwidth}
\centering
\includegraphics[width=0.9\textwidth]{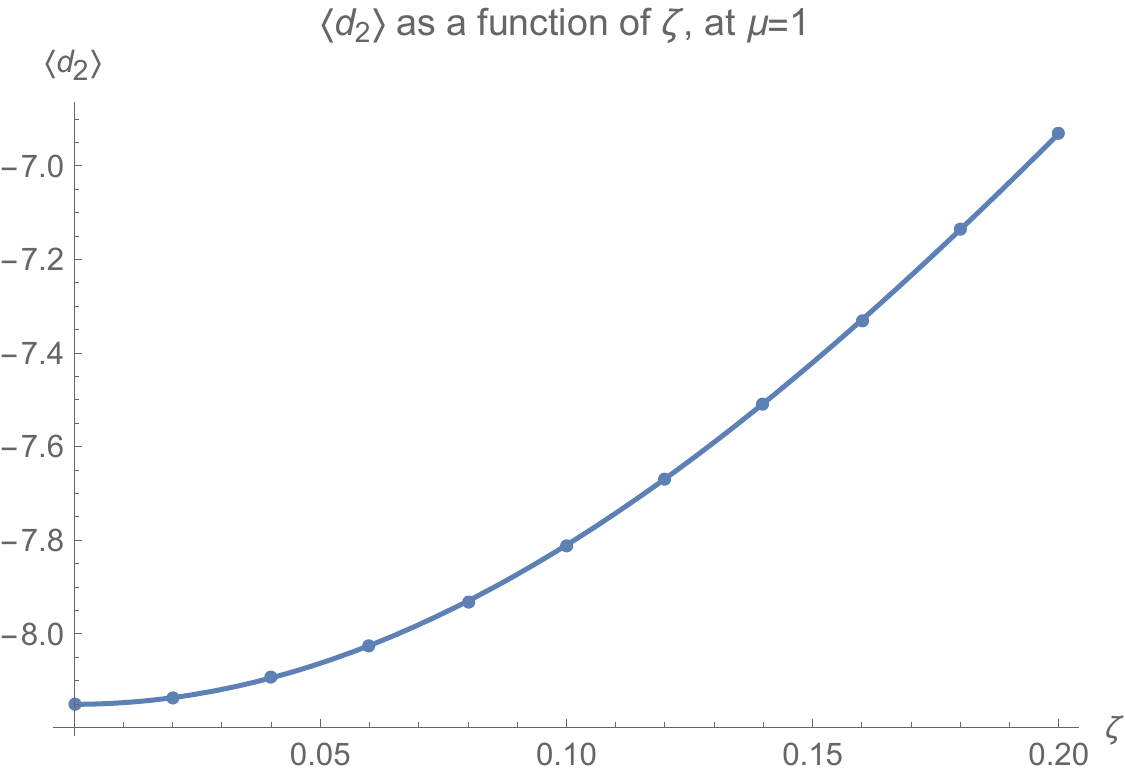}
\end{subfigure}%
\begin{subfigure}{0.5\textwidth}
\centering
\includegraphics[width=0.9\textwidth]{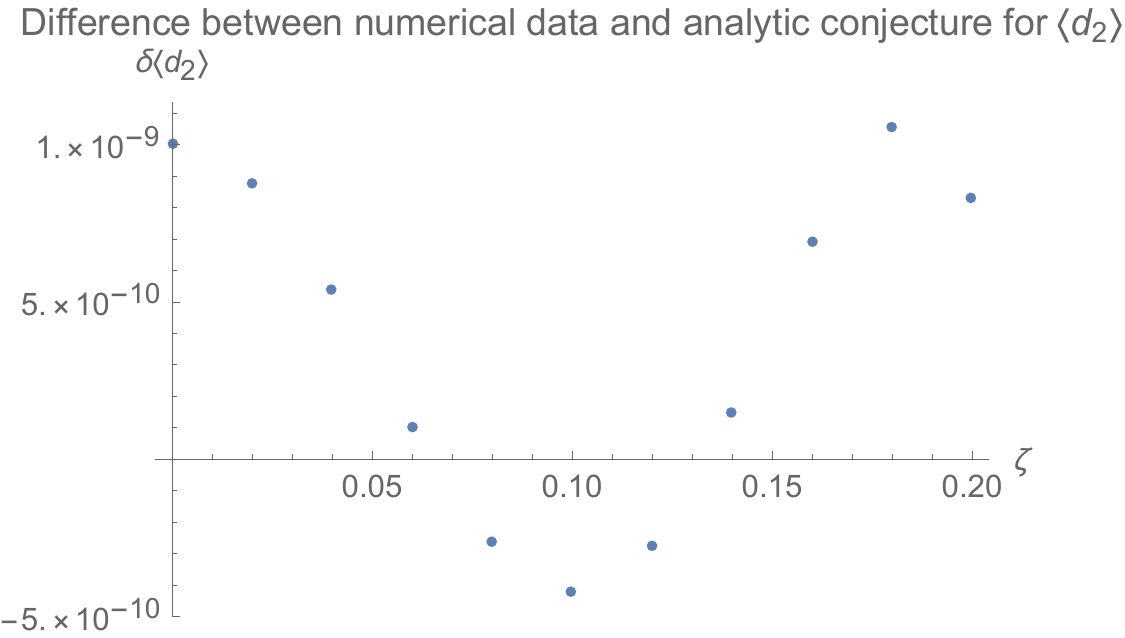}
\end{subfigure}
\caption{Numerical data compared with analytic conjecture for $\langle d_2\rangle$.}
\end{figure}

\begin{figure}
\centering
\begin{subfigure}{0.5\textwidth}
\centering
\includegraphics[width=0.9\textwidth]{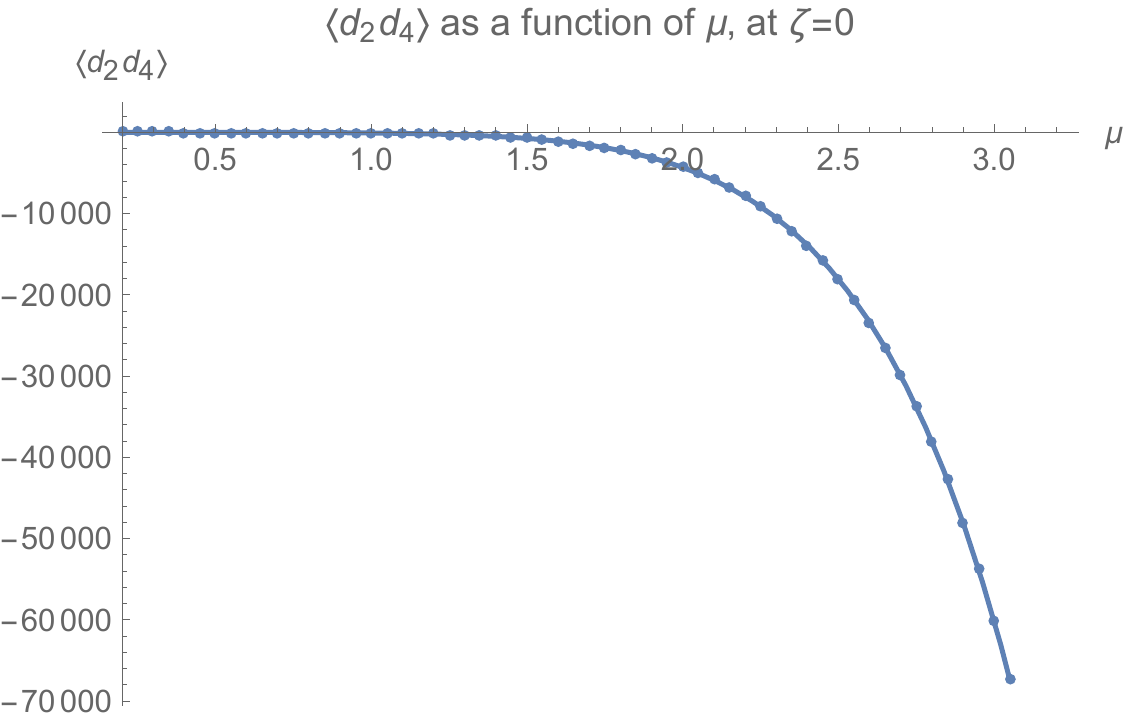}
\end{subfigure}%
\begin{subfigure}{0.5\textwidth}
\centering
\includegraphics[width=0.9\textwidth]{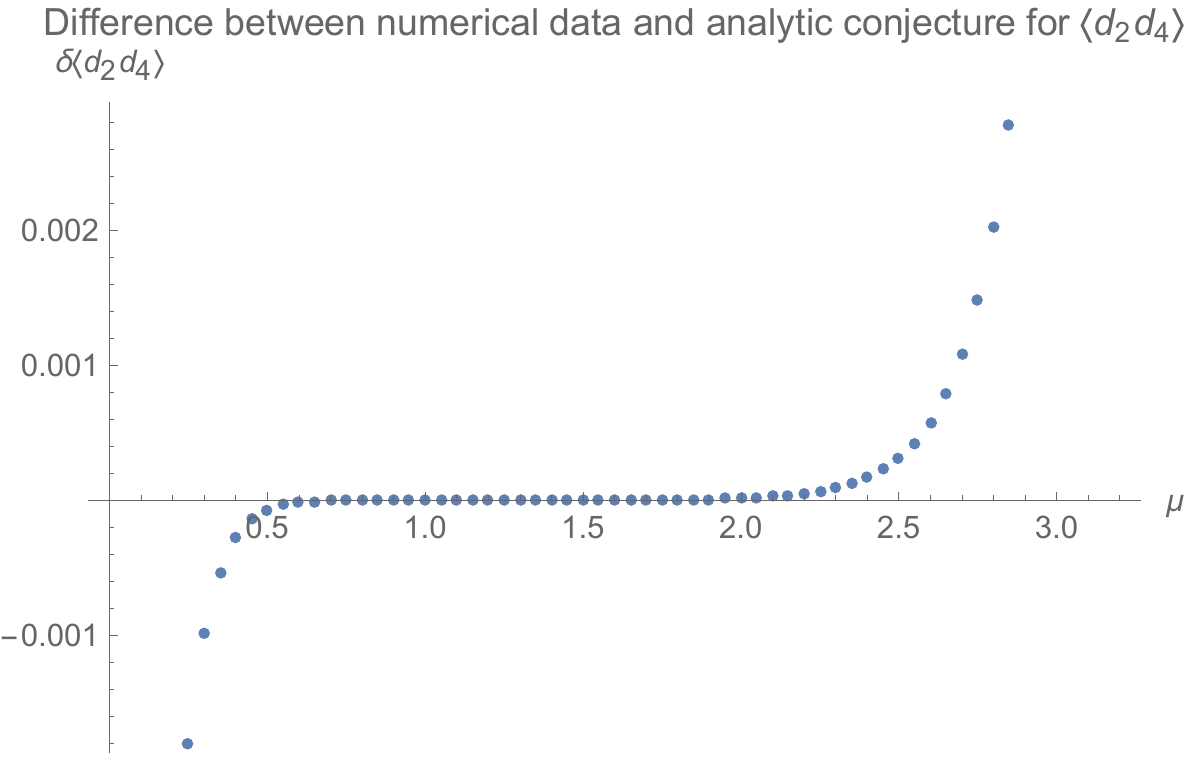}
\end{subfigure}
\begin{subfigure}{0.5\textwidth}
\centering
\includegraphics[width=0.9\textwidth]{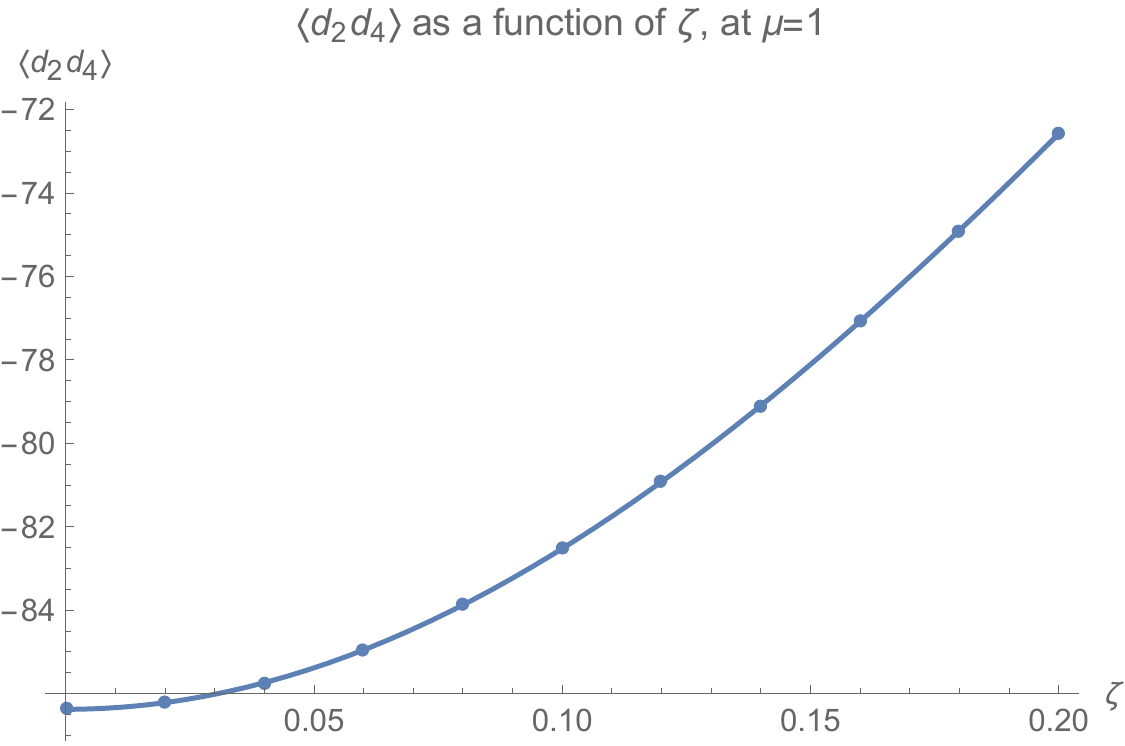}
\end{subfigure}%
\begin{subfigure}{0.5\textwidth}
\centering
\includegraphics[width=0.9\textwidth]{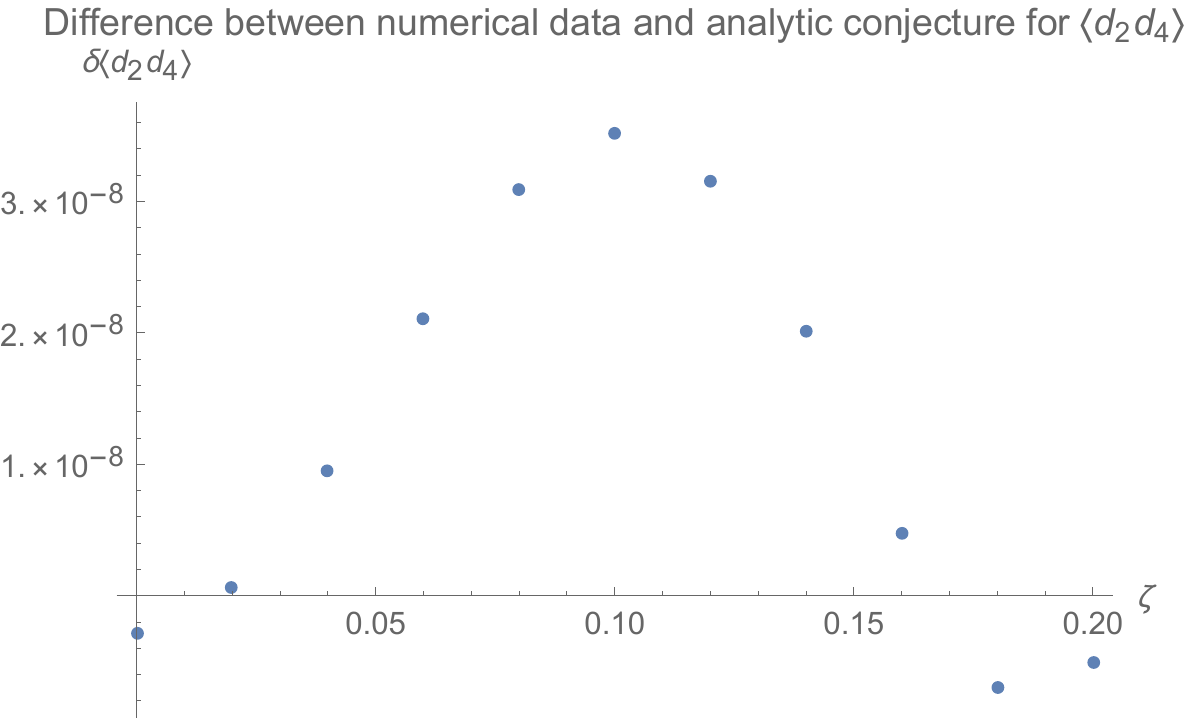}
\end{subfigure}
\caption{Numerical data compared with analytic conjecture for $\langle d_2 d_4\rangle$.}
\end{figure}

Similar tests were conducted for the case involving an $A_1$ singularity.  Good agreement was found between numerical results and our analytic conjecture for a range of values of $\mu$, $\zeta$, and $\zeta '$.  This agreement, for the cases of $\langle d_1 \rangle$ and $\langle d_2 \rangle$, are shown in Figures 3 and 4 below.

Due to the unavailability of an explicit presentation of the algebra $\aA^{(k+1)}$ for $k>1$, explicit analytic predictions for the perturbative correlation functions in the presence of an $A_k$ singularity are not currently at hand.  Nevertheless, the recursion relation satisfied by these correlation functions (equation \ref{eq:reck}) has been verified numerically for several one and two point functions.

\begin{figure}
\centering

\begin{subfigure}{0.5\textwidth}
\centering
\includegraphics[width=0.9\textwidth]{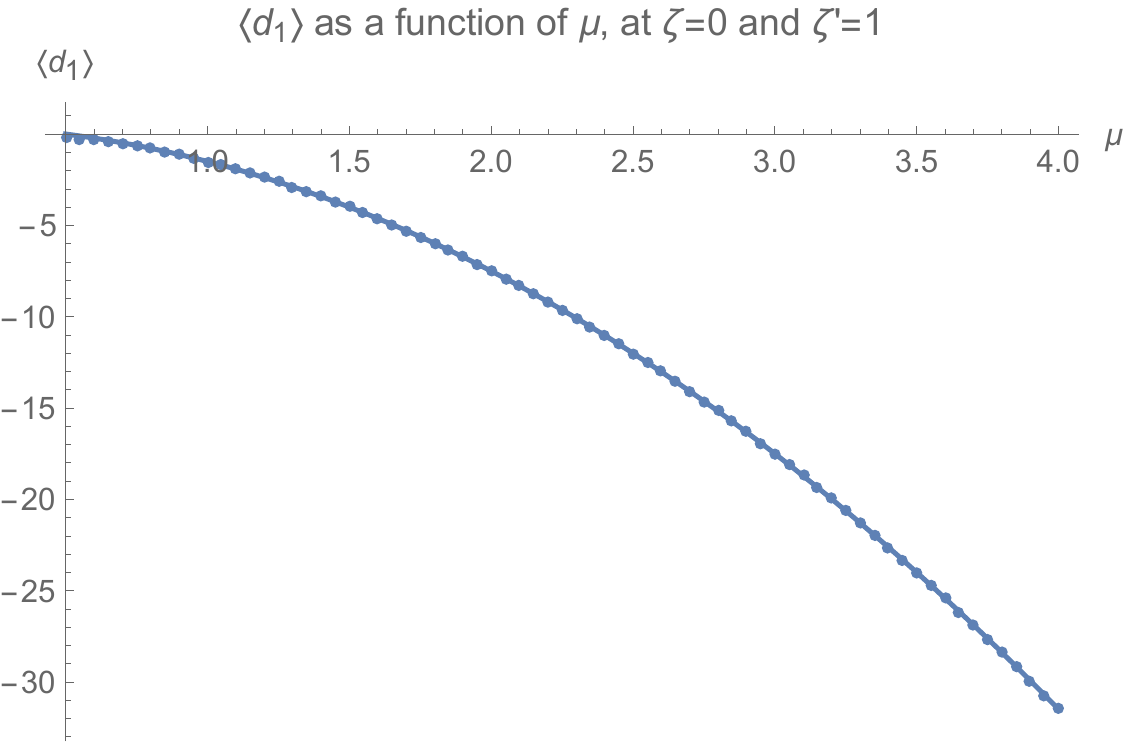}
\end{subfigure}%
\begin{subfigure}{0.5\textwidth}
\centering
\includegraphics[width=0.9\textwidth]{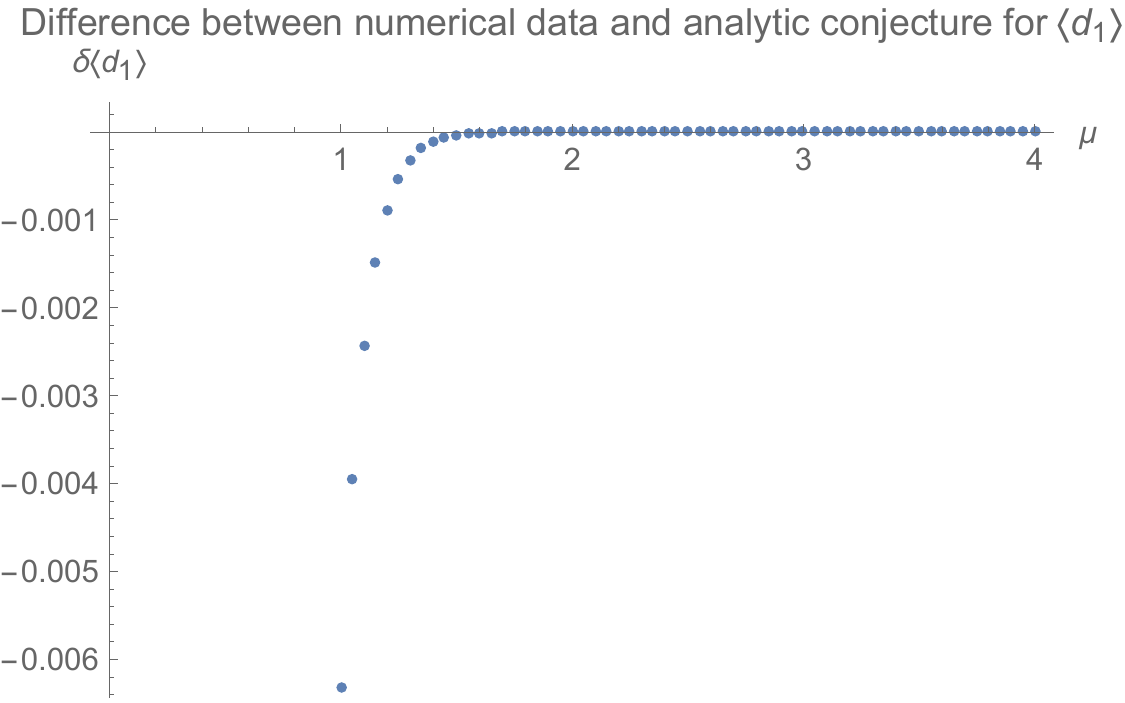}
\end{subfigure}

\begin{subfigure}{0.5\textwidth}
\centering
\includegraphics[width=0.9\textwidth]{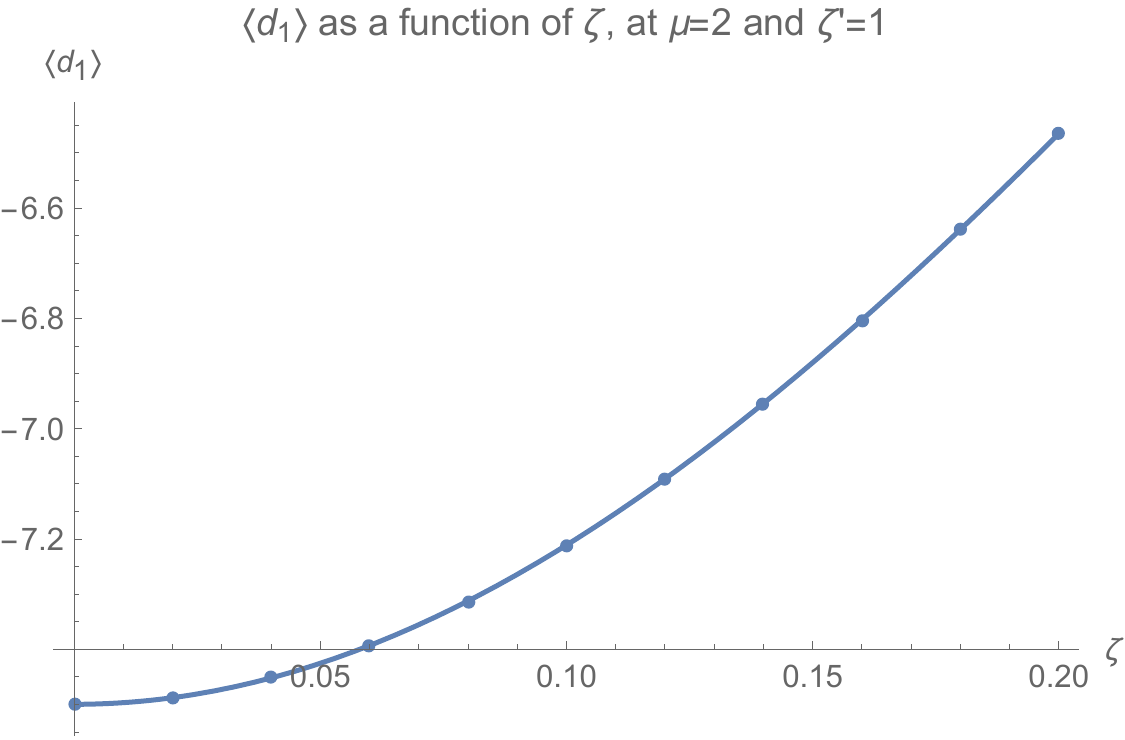}
\end{subfigure}%
\begin{subfigure}{0.5\textwidth}
\centering
\includegraphics[width=0.9\textwidth]{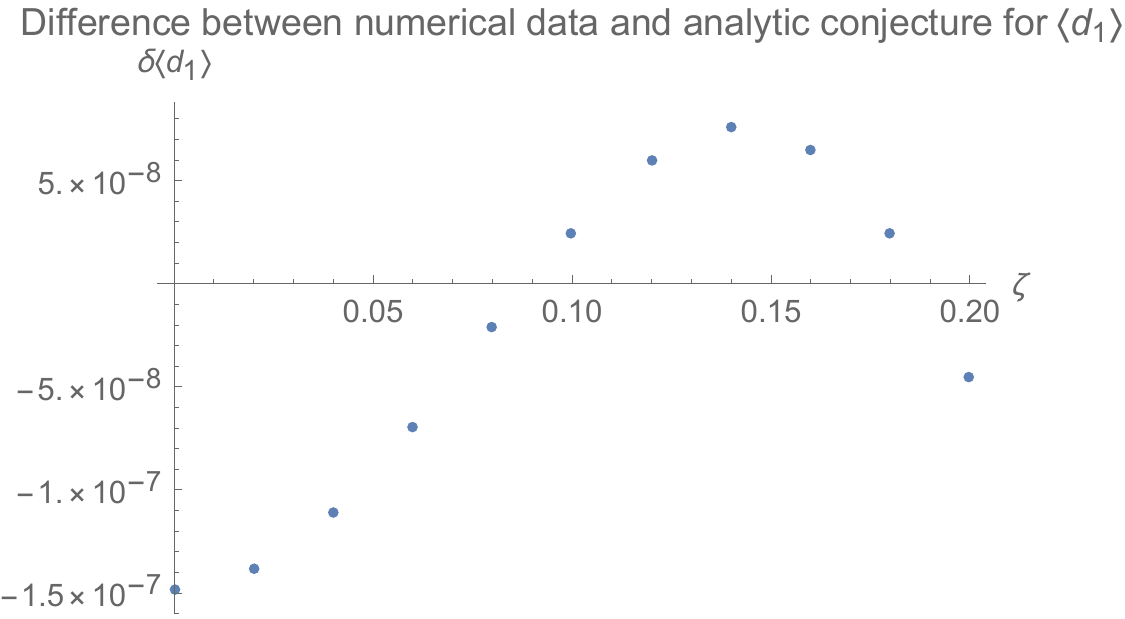}
\end{subfigure}

\begin{subfigure}{0.5\textwidth}
\centering
\includegraphics[width=0.9\textwidth]{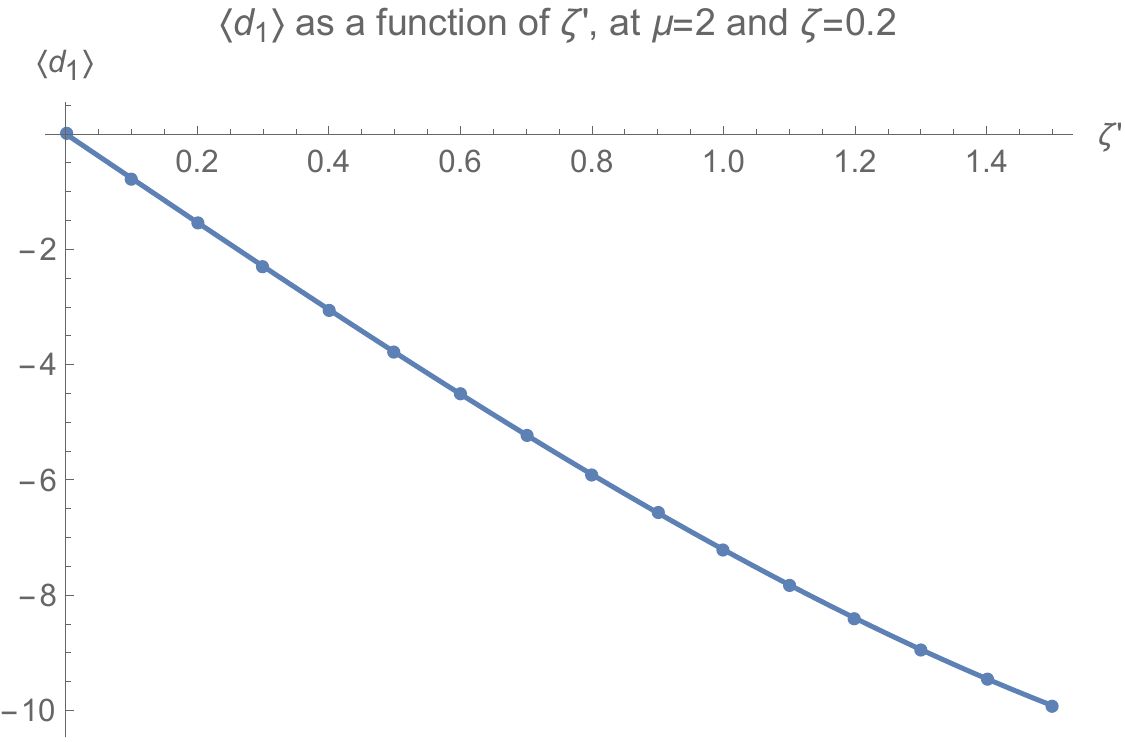}
\end{subfigure}%
\begin{subfigure}{0.5\textwidth}
\centering
\includegraphics[width=0.9\textwidth]{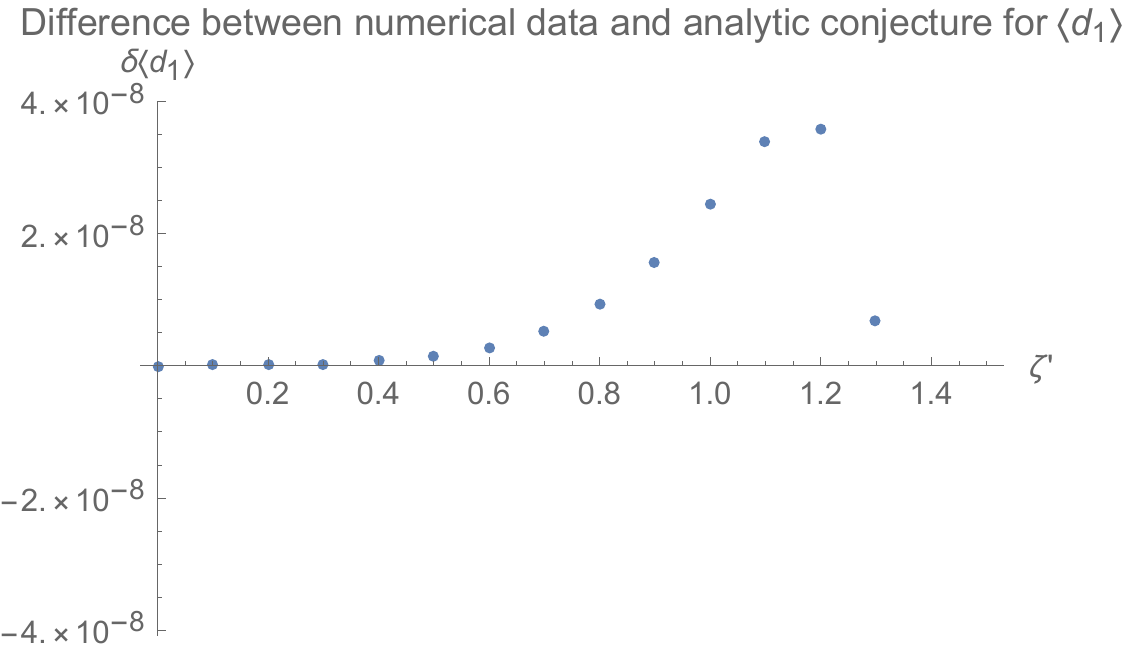}
\end{subfigure}
\caption{Numerical data compared with analytic conjecture for $\langle d_1\rangle$ in the presence of an $A_1$ singularity.}
\end{figure}

\begin{figure}
\centering

\begin{subfigure}{0.5\textwidth}
\centering
\includegraphics[width=0.9\textwidth]{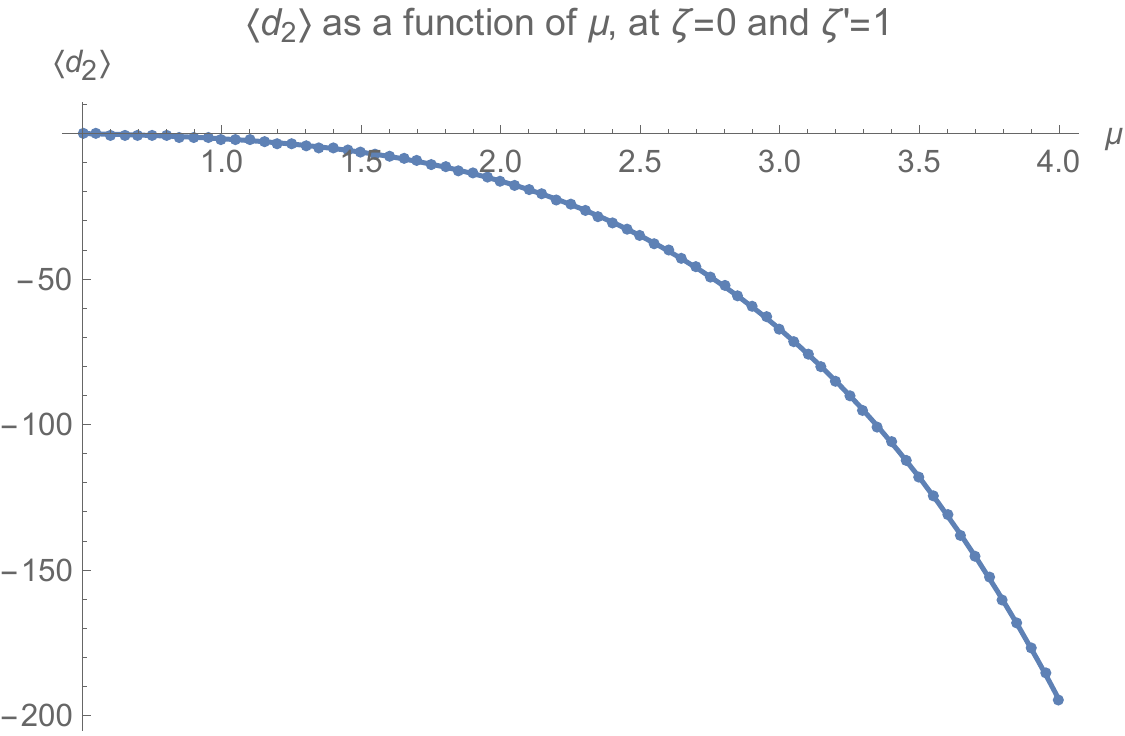}
\end{subfigure}%
\begin{subfigure}{0.5\textwidth}
\centering
\includegraphics[width=0.9\textwidth]{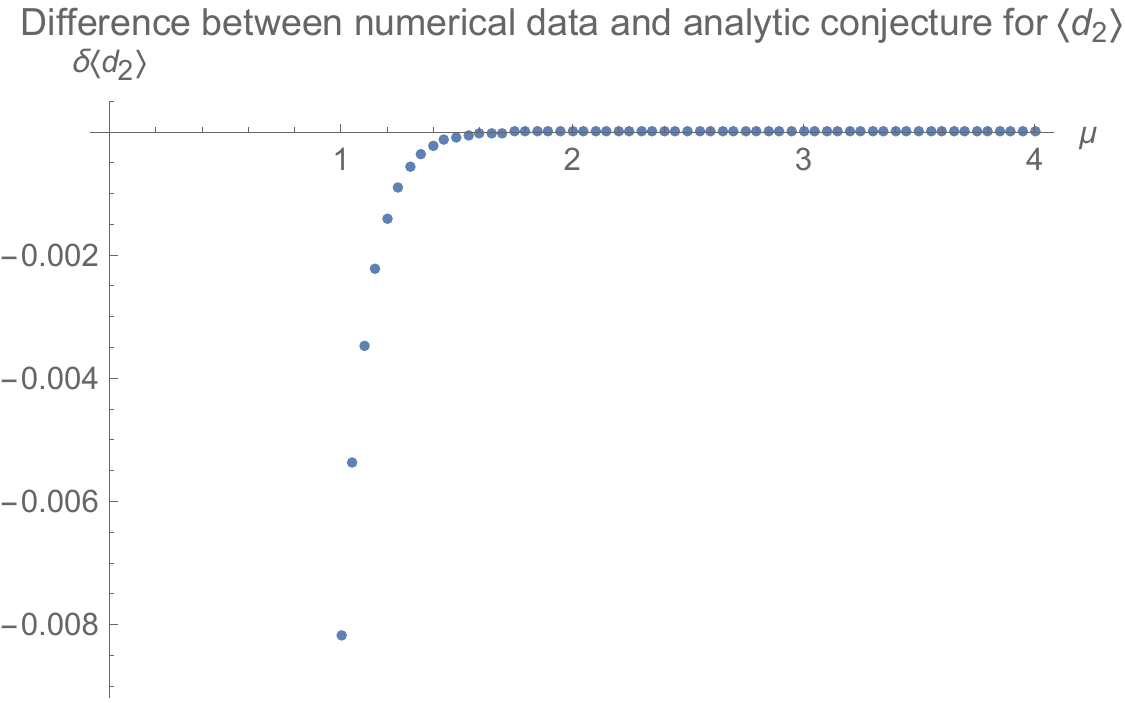}
\end{subfigure}

\begin{subfigure}{0.5\textwidth}
\centering
\includegraphics[width=0.9\textwidth]{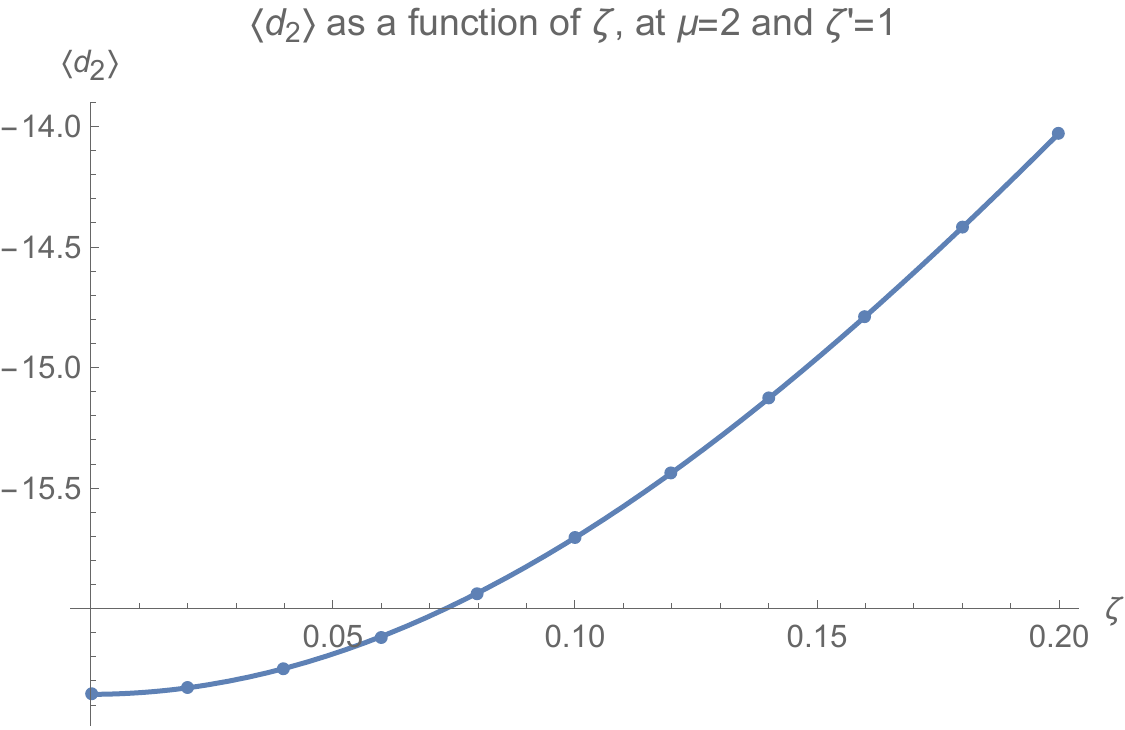}
\end{subfigure}%
\begin{subfigure}{0.5\textwidth}
\centering
\includegraphics[width=0.9\textwidth]{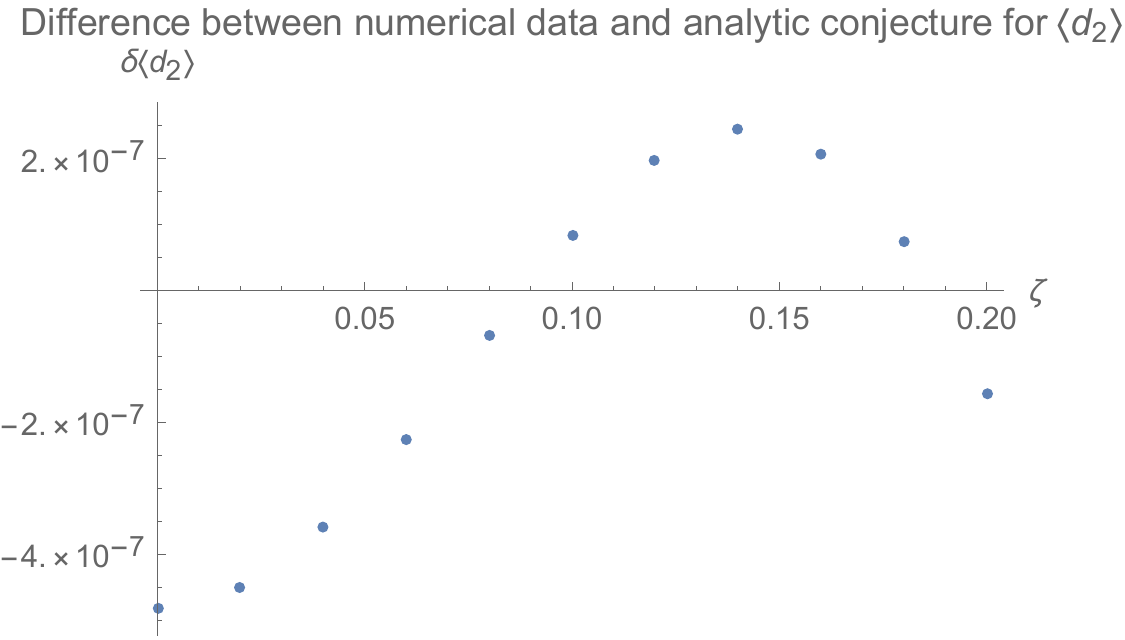}
\end{subfigure}

\begin{subfigure}{0.5\textwidth}
\centering
\includegraphics[width=0.9\textwidth]{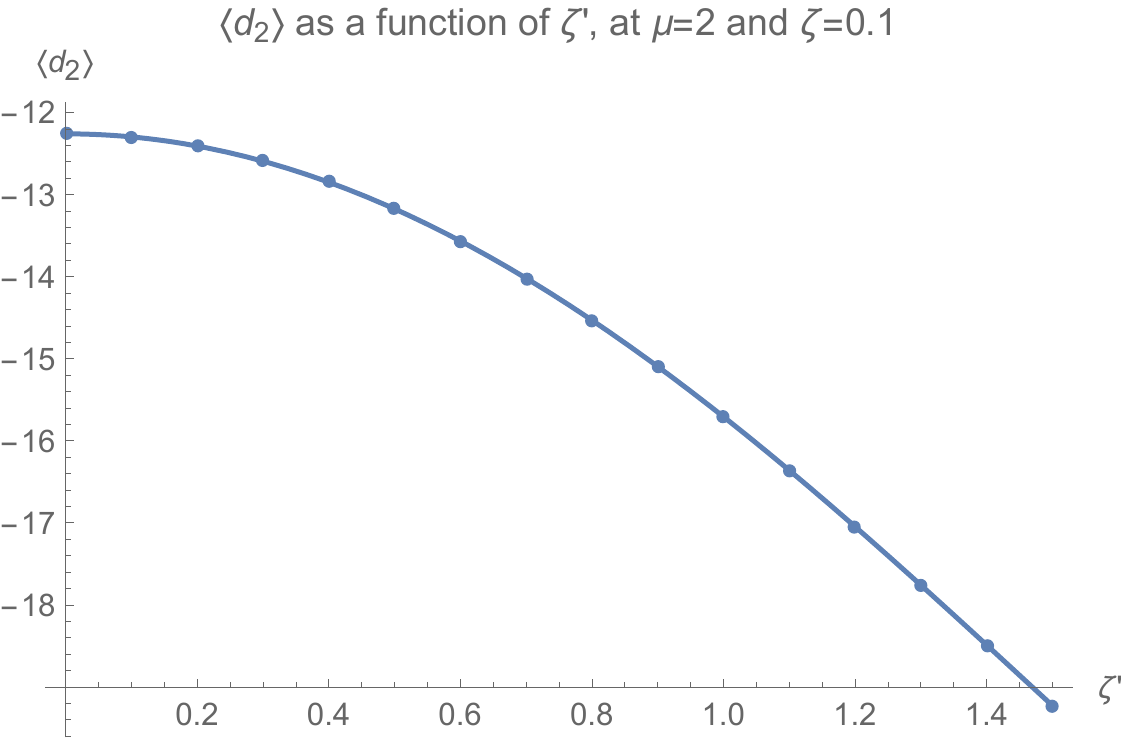}
\end{subfigure}%
\begin{subfigure}{0.5\textwidth}
\centering
\includegraphics[width=0.9\textwidth]{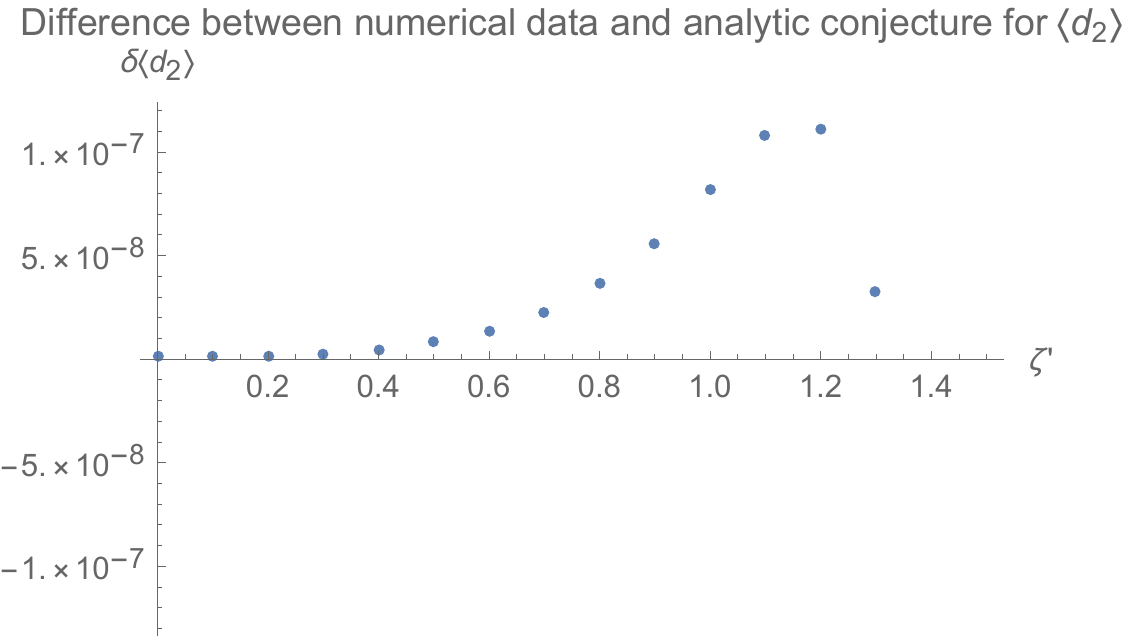}
\end{subfigure}
\caption{Numerical data compared with analytic conjecture for $\langle d_2\rangle$ in the presence of an $A_1$ singularity.}
\end{figure}

\clearpage
\bibliographystyle{JHEP}

\bibliography{mono}

\end{document}